\documentclass[reprint,aps]{revtex4-2}

\usepackage{physics}
\usepackage{amssymb}
\usepackage{graphicx}
\usepackage{mathtools}
\usepackage{mathrsfs}
\usepackage{dcolumn}
\usepackage{bm}
\usepackage[
colorlinks=true,
linkcolor=blue,
citecolor=blue,
urlcolor=blue
]{hyperref}

\begin{document}


\title{Surpassing the currently achievable distance of quantum key distribution based on sending-or-not-sending approach}


\author{Georgi Bebrov}
\affiliation{Department of Communication Engineering, Technical University of Varna, Varna, Bulgaria 9000}


\date{\today}

\begin{abstract}
Protocols based on the \textit{sending-or-not-sending} (SNS) principle have been intensively studied in recent years and have been shown to enable the longest transmission distances in quantum key distribution (QKD). In this work, we propose a \textit{sending-or-not-sending} \textit{phase-matching} QKD protocol (SNS-PM-QKD) that improves tolerance to phase mismatch, thereby extending the achievable transmission distance. We present a security analysis of SNS-PM-QKD in the asymptotic (infinite-key) regime under collective attacks. The performance of the proposed protocol is compared with that of standard phase-matching QKD, theoretical SNS-type twin-field QKD protocols (SNS-TF-QKD), and an experimental SNS-TF-QKD operated over transmission distances of up to $1002$km. Our results show that SNS-PM-QKD achieves greater transmission distances than these existing protocols, highlighting its potential for long-distance quantum communication.
\end{abstract}


\maketitle

\section{Introduction}
\indent Quantum key distribution (QKD) allows spatially separated parties to establish a shared, secure cryptographic key via an untrusted quantum channel supplemented by a fully authenticated classical channel \cite{bb84}. Over the years, several QKD protocols have been proposed to address implementation loopholes and performance limitations \cite{ekert,b92,hwang,eff-bb84,mdi}. A central limitation for the deployment of QKD remains the restricted operational transmission distance imposed by optical loss, phase (or polarization) drift in the communication channel, and error rates arising from device imperfections. The achievable distance of a QKD protocol is determined by its key-rate dependence on channel transmissivity and its key-rate tolerance to errors. The best transmissivity scaling of the key rate follows a square-root law \cite{lucamarini,cal,pmqkd,lin-lutkenhaus}, whereas the achievable error-rate tolerance is typically on the order of tens of percent \cite{sns}. Square-root transmissivity scaling is achieved in the class of twin-field QKD (TF-QKD) protocols \cite{lucamarini,cal,pmqkd,lin-lutkenhaus,sns}. This follows from a single-photon interference of bipartite coherent state at a beam splitter. A representative of this class is the so-called phase-matching measurement-device-independent QKD protocol, hereafter referred to as phase-matching QKD (PM-QKD) \cite{pmqkd,lin-lutkenhaus}. It is characterized by a simplified key-generation mechanism, lacking randomization \cite{lin-lutkenhaus}, relative to the original TF-QKD. Notably, this simplification does not lead to a degradation of protocol security. In general, tolerance to errors can be achieved in two ways: either through a key-generation mechanism that is fully (or partially) independent of the error rate, or by suppressing error induction via an appropriate key-generation mechanism together with a suitable postselection procedure \cite{sns,sns-post}. A relatively high tolerance to errors is attained in the so-called \textit{sending-or-not-sending} QKD protocol (SNS-QKD), a variant of TF-QKD \cite{sns,aopp,zigzag,sns-post,sns-428,sns-509,sns-511,snsexp2,sns-exp}. This follows from the absence of any interference in the key-generation events and the imposition of certain postselection criteria; a fairly large tolerance to misalignment errors can thus be demonstrated. \\ 
\indent So far SNS-TF-QKD has achieved the longest distance range among fiber-based QKD schemes. In the asymptotic limit, theoretical works \cite{sns,sns-post,zigzag,aopp} secure key generation over distances of up to approximately $910$km, while experiments have demonstrated record-breaking secure transmission distances of up to about $1002$km \cite{sns-exp}. \\
\indent Since long-distance operation is a central objective of QKD, particularly for wide-area quantum networks, it is of paramount importance to further improve protocol design so as to extend the rate–distance reach while preserving security. Motivated by this challenge, we propose a new QKD protocol inspired by the \textit{sending-or-not-sending} approach, specifically tailored to improve transmission distance under realistic conditions. Here we construct a protocol in which the error rate is further reduced by introducing \textit{outcome} postselection and \textit{sending-or-not-sending} paradigm into a PM-QKD framework. In PM-QKD, single-photon interference between the coherent states prepared by the legitimate parties (\textit{Alice} and \textit{Bob}) is performed at an untrusted intermediate node \cite{lin-lutkenhaus}. The key-generation process is based on the interference outcome\textemdash constructive or destructive\textemdash corresponding respectively to correlations or anticorrelations between the coherent-state phases chosen by \textit{Alice} and \textit{Bob}, with the key bits encoded into these phases. In the protocol proposed here, each legitimate party prepares two coherent states (a signal state and a secondary state) with identical phases and decides whether or not to send them to the intermediate node. 
At the intermediate node, a preliminary interference stage between the signal (secondary) states prepared by \textit{Alice} and \textit{Bob} is carried out at a coupler prior to the final joint interference measurement. The coupler intrinsically induces a $\pi$ phase shift in one of the interfering coherent states. Hence two coherent states with identical phases and intensities (ideal, loss-only scenario) interfere destructively upon meeting at the coupler. Consequently, depending on the \textit{sending-or-not-sending} choice—namely, the \textit{sending–not sending} (\textit{not sending–sending}), \textit{sending–sending}, or \textit{not sending–not sending} scenario—the joint measurement registers either no click or a click only at the detector corresponding to destructive interference. In the ideal, loss-only scenario, a detection event occurs exclusively for the \textit{sending–not-sending} (or \textit{not-sending–sending}) configuration. As in the original SNS-QKD protocol, key generation in the proposed scheme is also based on the \textit{sending-or-not-sending} choice. Under realistic conditions, relevant detection events occur in all \textit{sending-or-not-sending} configurations due to dark counts and misalignment in both the couplers and the final interference beam splitter. A characteristic feature of the proposed QKD model is that sifting all measurement outcomes except the conclusive detection corresponding to destructive interference (outcome postselection) mitigates the error rate. This mitigation leads to an improved key-rate\textendash to\textendash error-rate ratio, which in turn enables greater operational distances. \\
\indent For security analysis we adopt the framework introduced in Ref. \cite{lin-lutkenhaus}. To evaluate the performance of the newly proposed protocol, we compare it to PM-QKD \cite{lin-lutkenhaus}, theoretical SNS-TF-QKD models \cite{sns,sns-post}, and experimental SNS-TF-QKD implementation \cite{sns-exp} in terms of their achievable transmission (operational) distances. Refs.~\cite{sns-post} and \cite{sns-exp} report the maximum achievable distances to date for theoretical and experimental \textit{sending-or-not-sending} protocols, respectively. \\
\indent The remainder of this paper is organized as follows. In Sec. \ref{protocol}, we present the detailed description of the proposed protocol. In Sec. \ref{security}, we provide the security analysis. Numerical simulations and performance results are reported in Sec. \ref{sim}. Finally, Sec. \ref{summ} summarizes the main conclusions and discusses possible directions for future work.

\section{Proposed protocol}\label{protocol}

\subsection{Sending-or-not-sending phase-matching QKD}\label{sns-pm-qkd}
\begin{figure}[!hbt]
	\centering
	\includegraphics[scale=0.35]{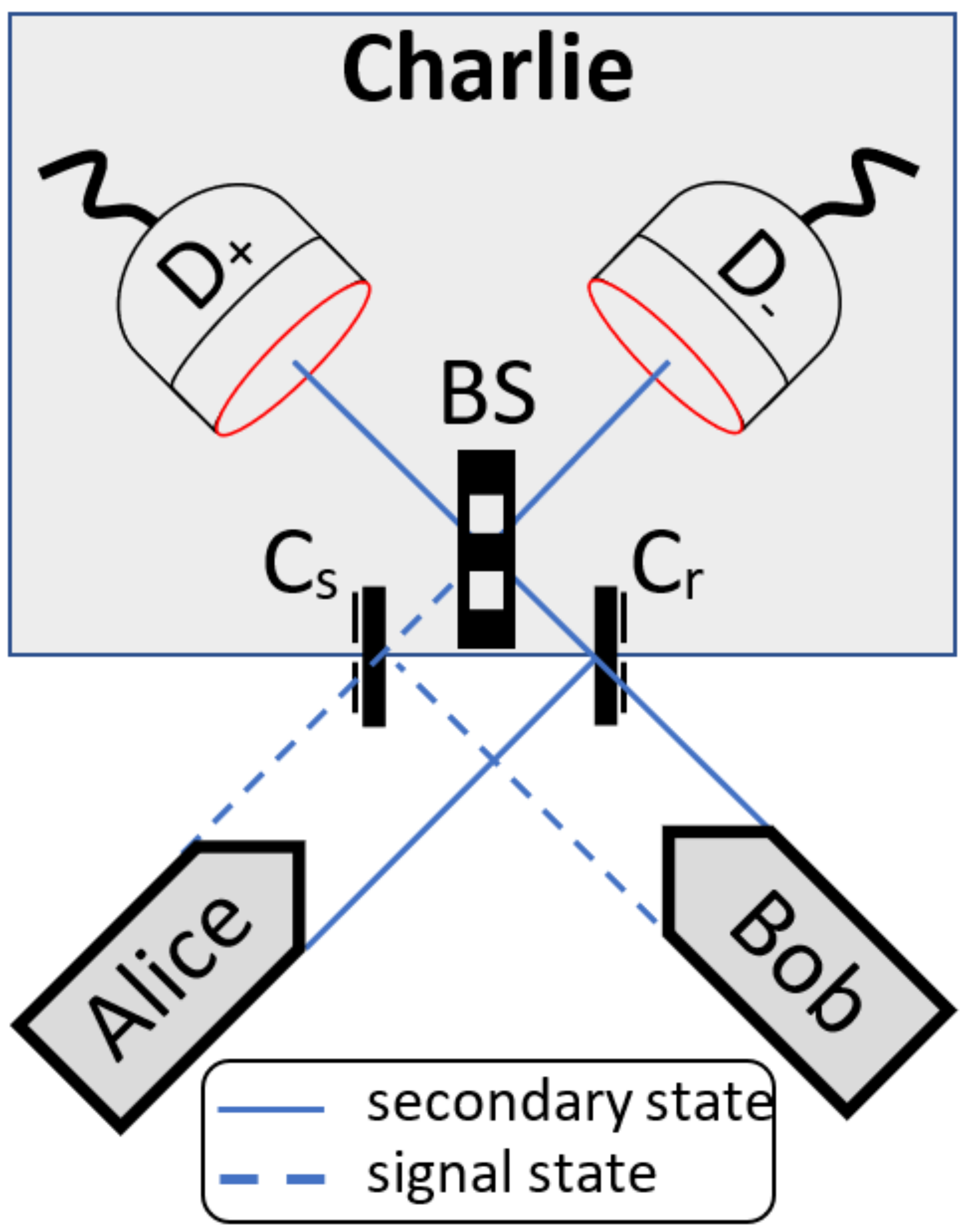}
	\caption{\label{setup} Schematic setup of SNS-PM-QKD.
		\textit{Alice} (\textit{Bob}) independently chooses to send or not to send \textit{her} (\textit{his}) signal and secondary coherent states to a third party \textit{Charlie}. \textit{He} performs a \textit{joint} (\textit{beam splitter}) \textit{measurement} on the coherent states coming from the couplers, \textbf{C}$_{\text{\textbf{s}}}$ and \textbf{C}$_{\text{\textbf{r}}}$. The couplers output: either \textbf{1)} the interference of \textit{Alice}'s and \textit{Bob}'s signal states (output of \textbf{C}$_{\text{\textbf{s}}}$) and interference of \textit{Alice}'s and \textit{Bob}'s secondary (reference) states (output of \textbf{C}$_{\text{\textbf{r}}}$)\textemdash \textit{sending-sending} \textbf{ss} scenario; or \textbf{2)} \textit{Alice}'s signal state (output of \textbf{C}$_{\text{\textbf{s}}}$) and \textit{Alice's} secondary state (output of \textbf{C}$_{\text{\textbf{r}}}$)\textemdash \textit{sending-not sending} (\textbf{sns}) scenario; or \textbf{3)} \textit{Bob}'s signal state (output of \textbf{C}$_{\text{\textbf{s}}}$) and \textit{Bob}'s secondary state (output of \textbf{C}$_{\text{\textbf{r}}}$)\textemdash \textit{not sending-sending} (\textbf{sns}) scenario; or \textbf{4)} nothing (neither \textbf{C}$_{\text{\textbf{s}}}$ nor \textbf{C}$_{\text{\textbf{r}}}$ outputs a coherent state)\textemdash \textit{not sending-not sending} (\textbf{nn}) scenario. Next \textit{Charlie} publicly announces the measurement outcome. \textbf{BS}\textemdash balanced beam splitter; \textbf{D}\textemdash single-photon detector; \textbf{C}\textemdash coupler.}
\end{figure}

\indent In this section, we propose a protocol based on the setup of Fig. \ref{setup}. The protocol is called \textit{sending-or-not-sending} \textit{phase-matching} \textit{quantum key distribution} (SNS-PM-QKD). It consists of the following steps: \\
\textit{Step 1}. \textit{Alice} (\textit{Bob}) generates a random number $m_a$ ($m_b$) $\in \{0,1\}$ with respect to a probability distribution $\{p_a, 1-p_a\}$ ($\{p_b, 1-p_b\}$) ~\cite{lin-lutkenhaus}. If \textit{Alice} and \textit{Bob} choose $m_a = 0$ and $m_b = 0$, respectively, the current protocol round is considered as a \textit{signal} round. If \textit{Alice} and Bob \textit{choose} $m_a = 1$ and $m_b = 1$, respectively, the round is considered as a \textit{test} round.\\
\textit{Step 2}. \textit{State preparation}\textemdash In case of a test round, \textit{Alice} (\textit{Bob}) chooses whether or not to send coherent states (both test and secondary (or \textit{reference}) states) to \textit{Charlie} with probability $\epsilon$. Several scenarios then take place when the approach of \textit{sending}-\textit{not}-\textit{sending} is applied~\cite{sns}: sending-not sending \textit{or} not sending-sending (\textbf{sns}), not sending-not sending (\textbf{nn}), and sending-sending (\textbf{ss}). If \textit{Alice} (\textit{Bob}) chooses to send coherent states to \textit{Charlie}, \textit{she} (\textit{he}) prepares a test coherent state $\ket{\mathrm{e}^{i\omega_a}\sqrt{\mu_a}}$ $\left(\ket{\mathrm{e}^{i\omega_b}\sqrt{\mu_b}}\right)$ and a secondary (or \textit{reference}) coherent state $\ket{\mathrm{e}^{i\theta_a}\sqrt{\mu_a}}$ $\left(\ket{\mathrm{e}^{i\theta_b}\sqrt{\mu_b}}\right)$, where $\theta_{a(b)}\in[0,2\pi)$, $\omega_{a(b)}\in[0,2\pi)$, and $\mu_{a(b)}$ are chosen at random. Thus a test round configuration ($\mu_{a(b)}$, $\theta_{a(b)}$, $\omega_{a(b)}$) of this protocol is identical to the configuration utilized in Ref. ~\cite{lin-lutkenhaus}. Note that in the test rounds, \textit{Alice} and \textit{Bob} announce their \textit{sending-or-not-sending} choices, $\mu_{a(b)}$ choices, $\omega_{a(b)}$ and $\theta_{a(b)}$ choices. \\
\indent In case of a signal round, \textit{Alice} (\textit{Bob}) chooses whether or not to send coherent states (both signal and secondary) to \textit{Charlie} with probability $\epsilon$. Several scenarios then take place when the approach of \textit{sending}-\textit{or}-\textit{not}-\textit{sending} is applied~\cite{sns}: sending-not sending \textit{or} not sending-sending (\textbf{sns}), not sending-not sending (\textbf{nn}), and sending-sending (\textbf{ss}). If \textit{Alice} (\textit{Bob}) chooses to send coherent states to \textit{Charlie}, \textit{she} (\textit{he}) prepares a signal coherent state $\ket{\mathrm{e}^{i\varphi_a}\sqrt{\mu}}_{Q}$ $\left(\ket{\mathrm{e}^{i\varphi_b}\sqrt{\mu}}_{Q'}\right)$ and a secondary coherent state $\ket{\mathrm{e}^{i\tau_a}\sqrt{\mu}}_{T}$ $\left(\ket{\mathrm{e}^{i\tau_b}\sqrt{\mu}}_{T'}\right)$, where $\tau_a=0$, $\tau_b=0$, $\varphi_a=0$, and $\varphi_b=0$. Note that, in case of a noiseless (no-phase-drift) quantum channel, if both parties choose to send coherent states, their states cancel each other at the couplers \textbf{C}$_{\text{\textbf{s}}}$, \textbf{C}$_{\text{\textbf{r}}}$\textemdash the secondary states of \textit{Alice} and \textit{Bob} interfere destructively at \textbf{C}$_{\text{\textbf{r}}}$, while their signal states interfere destructively at \textbf{C}$_{\text{\textbf{s}}}$. Note further that, in case of a noiseless quantum channel, when \textbf{sns} scenario occurs, only detector $D_-$ should click ($\Delta\phi_{a(b)} = \abs{\varphi_{a(b)} - \tau_{a(b)}} = \pi$)~\footnote{\textit{Note}: Phases $\varphi_b$ and $\tau_a$ picks an additional value of $\pi$ ($\varphi_b=0+\pi$, $\tau_a = 0 + \pi$) at couplers \textbf{C}$_{\text{\textbf{s}}}$ and \textbf{C}$_{\text{\textbf{r}}}$, respectively, since the corresponding coherent states $\ket{\mathrm{e}^{i\varphi_b}\sqrt{\mu}}_{Q'}$ and $\ket{\mathrm{e}^{i\tau_a}\sqrt{\mu}}_{T}$ undergo reflection at these devices.}; see Fig. \ref{setup}. \\
\textit{Step 3}. \textit{Measurement}\textemdash \textit{Alice} and \textit{Bob} send their coherent states (signal/test and secondary states) to the intermediate node \textit{Charlie}. At the entry of \textit{Charlie}'s side, in both test and signal rounds, the signal/test coherent states of \textit{Alice} and \textit{Bob} are subjected to an interference at coupler \textbf{C}$_{\text{\textbf{s}}}$, whereas the secondary coherent states of \textit{Alice} and \textit{Bob} are subjected to an interference at coupler \textbf{C}$_{\text{\textbf{r}}}$: the output state of \textbf{C}$_{\text{\textbf{s}}}$ is $\ket{\mathrm{e}^{i\varphi_a}\sqrt{\mu} - \mathrm{e}^{i\varphi_b}\sqrt{\mu}}$; the output state of \textbf{C}$_{\text{\textbf{r}}}$ is $\ket{\mathrm{e}^{i\tau_b}\sqrt{\mu} - \mathrm{e}^{i\tau_a}\sqrt{\mu}}$. The interference process taken place at a coupler is described in Appendix \ref{coupler}. The outputs of these couplers (results of the interference processes just mentioned) are then directed towards a beamsplitter \textbf{BS}. For each round, \textit{Charlie} performs a joint (beamsplitter) measurement that is characterized by POVM elements $F^{\varkappa}$ ~\cite{lin-lutkenhaus}. He announces the measurement outcome $\varkappa \in \{+,-,?,d\}$, where "$+$" identifies the presence of a click only at detector $D_+$, "$-$" identifies the presence of a click only at detector $D_-$, "$?$" identifies the absence of a detector click, and "$d$" identifies the presence of clicks at both detectors ~\cite{lin-lutkenhaus}; see Fig. \ref{setup}.
\begin{center}
	\textit{Steps 1\textendash3 are iterated $N$ times ($N\rightarrow\infty$), an asymptotic regime of operation is considered}.
\end{center}
\textit{Step 4}. \textit{Public communication and sifting}\textemdash A public disclosure of information over an authenticated public classical channel takes place in this step. \textit{Alice} and \textit{Bob} first announce their values of $m_a$, $m_b$\textemdash\textit{they} perform \textit{round-type} sifting. In case of a common \textit{signal} round ($m_{a(b)} = 0$), both \textit{Alice} and \textit{Bob} also performs \textit{outcome sifting}\textemdash\textit{they} keep a common signal round ($m_{a}=m_b=0$) \textit{if and only if} a measurement outcome $\varkappa=-$ is registered and consequently announced by \textit{Charlie}. Note that the signal rounds, retained after the outcome sifting, are used to distill secret-key bits.  \\
\indent In case of a common \textit{test} round ($m_{a(b)} = 1$), \textit{Alice} and \textit{Bob} announce their intensities $\mu_{a(b)}$, \textit{their} phases $\theta_{a(b)}$ and $\omega_{a(b)}$, and their choices of \textit{sending} or \textit{not sending} (\textit{Alice} and \textit{Bob} thus determines the scenario of this round\textemdash \textbf{sns}, \textbf{nn}, or \textbf{ss} scenario). \textit{Alice} and \textit{Bob} further perform \textit{scenario sifting} (only \textbf{sns} scenarios are kept), as mentioned earlier. All the test rounds, retained after scenario sifting, are involved in the parameter estimation procedure.\\
\textit{Step 5}. \textit{Parameter estimation}\textemdash Here we adopt the \textit{quantum channel tomography} ~\cite{lin-lutkenhaus} to evaluate the influence of an eavesdropper on both $\ket{\mathrm{e}^{i\varphi_a}\sqrt{\mu},\mathrm{e}^{i\tau_a}\sqrt{\mu}}_{QT}$ and $\ket{\mathrm{e}^{i\varphi_b}\sqrt{\mu},\mathrm{e}^{i\tau_b}\sqrt{\mu}}_{Q'T'}$ coherent-state pairs. Based on \textit{Step 3} and \textit{Step 4} announcements related to test rounds, \textit{Alice} and Bob estimate the information an eavesdropper (\textit{Eve}) gains out of the coherent states being transferred in the protocol. If \textit{Eve} gains information exceeding a certain threshold, the case in which no secret key can be established, \textit{Alice} and \textit{Bob} terminate the protocol ~\cite{lin-lutkenhaus}. Otherwise, they proceed to the next steps. \\
\textit{Step 6}. \textit{Establishing key bits}\textemdash Given a common signal round and announced outcome $\varkappa=-$, \textit{Alice} and \textit{Bob} establish \textit{their} raw key bits $k$ and $y$, respectively, according to the following expressions
\begin{equation}
	k = \begin{cases}
		0,\; \text{\textit{sending}}\\
		1,\; \text{\textit{not sending}} \\
	\end{cases}; \;\;\;
	y = \begin{cases}
		0,\; \text{\textit{sending}} \\
		1,\; \text{\textit{not sending}}\\
	\end{cases}.
\end{equation}
That is, $k=\overline{y}$ is required in order for \textit{Alice} and \textit{Bob} to share correlated key bits. Note that announced outcome $\varkappa=-$ signifies the Boolean inversion relation between $k$ and $y$\textemdash Bob inverts his key bit $y$ given $\varkappa=-$ takes place ($y$ $\rightarrow$ $\overline{y}$). \\
\textit{Step 7}. \textit{Alice} and \textit{Bob} perform error correction and privacy amplification to their raw keys in order to distill a common secret key. 

\subsection{SNS-PM-QKD with phase randomization}\label{sns-pm-qkd-r}
\begin{figure}[!hbt]
	\centering
	\includegraphics[scale=0.35]{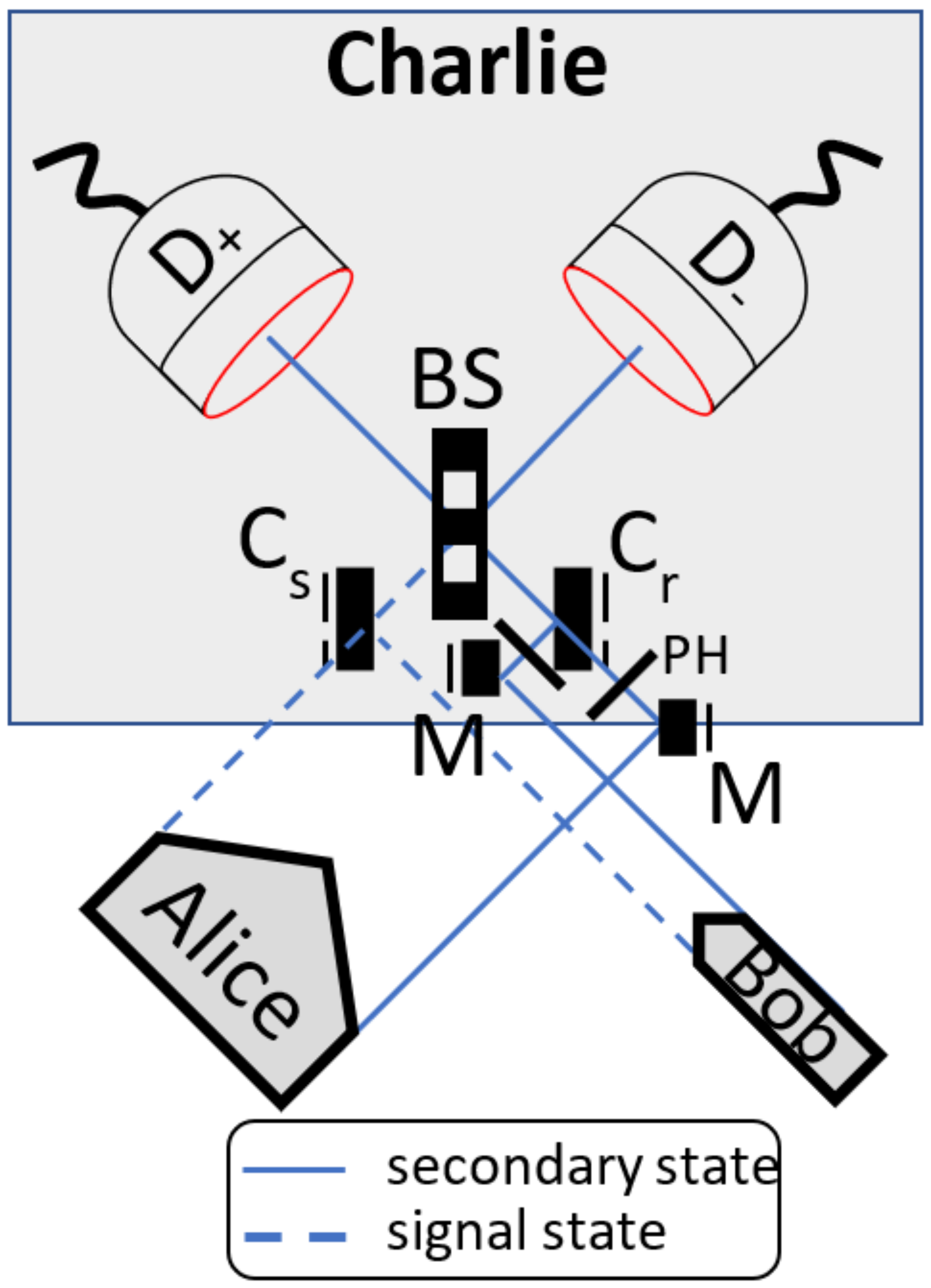}
	\caption{\label{setup-rand} Schematic setup of SNS-PM-QKD with phase randomization. \textit{Alice} (\textit{Bob}) independently chooses to send or not to send \textit{her} (\textit{his}) signal and secondary coherent states to a third party, \textit{Charlie}. \textit{He} performs a \textit{joint} (\textit{beam splitter}) \textit{measurement} on the coherent states coming from the couplers, \textbf{C}$_{\text{\textbf{s}}}$ and \textbf{C}$_{\text{\textbf{r}}}$. The couplers output: either \textbf{1)} the interference of \textit{Alice}'s and \textit{Bob}'s signal states (output of \textbf{C}$_{\text{\textbf{s}}}$) and interference of \textit{Alice}'s and \textit{Bob}'s secondary (reference) states (output of \textbf{C}$_{\text{\textbf{r}}}$)\textemdash \textit{sending-sending} \textbf{ss} scenario; or \textbf{2)} \textit{Alice}'s signal state (output of \textbf{C}$_{\text{\textbf{s}}}$) and \textit{Alice's} secondary state (output of \textbf{C}$_{\text{\textbf{r}}}$)\textemdash \textit{sending-not sending} (\textbf{sns}) scenario; or \textbf{3)} \textit{Bob}'s signal state (output of \textbf{C}$_{\text{\textbf{s}}}$) and \textit{Bob}'s secondary state (output of \textbf{C}$_{\text{\textbf{r}}}$)\textemdash \textit{not sending-sending} (\textbf{sns}) scenario; or \textbf{4)} nothing (neither \textbf{C}$_{\text{\textbf{s}}}$ nor \textbf{C}$_{\text{\textbf{r}}}$ outputs a coherent state)\textemdash \textit{not sending-not sending} (\textbf{nn}) scenario. In the above schematic, the secondary states of \textit{Alice} and \textit{Bob} swap positions by means of mirrors \textbf{M} and phase shifts \textbf{PH} of $\pi$ [rad]. Note that the role of a \textbf{PH} is to cancel the phase shift induced into a coherent state upon its reflection off a mirror. After a conclusive measurement, \textit{Charlie} publicly announces the corresponding outcome. \textbf{BS}\textemdash balanced beam splitter; \textbf{D}\textemdash single-photon detector; \textbf{C}$_{\text{\textbf{x}}}$\textemdash coupler (\textbf{x} $=$ r\textemdash coupler for secondary (reference) states; \textbf{x} $=$ s\textemdash coupler for signal states); \textbf{M}\textemdash mirror; \textbf{PH}\textemdash phase shift of $\pi$ [rad].}
\end{figure}

\indent To be consistent with the SNS-QKD framework \cite{sns,sns-post}, a SNS-PM-QKD protocol with \textit{phase randomization} is introduced below. The protocol is based on a setup depicted in Fig. \ref{setup-rand}, a slight modification of Fig. \ref{setup}. It consists of the following steps. \\
\textit{Step 1 $'$}. This step is identical to \textit{Step 1}. \\
\textit{Step 2 $'$}. \textit{State preparation}\textemdash In case of a test round, \textit{Alice} (\textit{Bob}) chooses whether or not to send coherent states (both test and secondary (or \textit{reference})) to \textit{Charlie} with probability $\epsilon$. If \textit{Alice} (\textit{Bob}) chooses to send coherent states to \textit{Charlie}, \textit{she} (\textit{he}) prepares a test coherent state $\ket{\mathrm{e}^{i\omega_a}\sqrt{\mu_a}}$ $\left(\ket{\mathrm{e}^{i\omega_b}\sqrt{\mu_b}}\right)$ and a secondary (or \textit{reference}) coherent state $\ket{\mathrm{e}^{i\theta_a}\sqrt{\mu_a}}$ $\left(\ket{\mathrm{e}^{i\theta_b}\sqrt{\mu_b}}\right)$, where $\theta_{a(b)}\in[0,2\pi)$, $\omega_{a(b)}\in[0,2\pi)$, and $\mu_{a(b)}$ are chosen at random. \\
In case of a signal round, \textit{Alice} (\textit{Bob}) chooses whether or not to send coherent states (both signal and secondary) to \textit{Charlie} with probability $\epsilon$. If \textit{Alice} (\textit{Bob}) chooses to send coherent states to Charlie, \textit{she} (\textit{he}) prepares a signal coherent state $\ket{\mathrm{e}^{i\nu_a}\sqrt{\mu}}_{Q}$ $\left(\ket{\mathrm{e}^{i\nu_b}\sqrt{\mu}}_{Q'}\right)$ and a secondary (or \textit{reference}) coherent state $\ket{\mathrm{e}^{i\nu_a}\sqrt{\mu}}_{T}$ $\left(\ket{\mathrm{e}^{i\nu_b}\sqrt{\mu}}_{T'}\right)$, where $\nu_a,\nu_b\in[0,2\pi)]$ are chosen randomly and independently of each other. The choice of $\nu_a$ ($\nu_b$) is as follows. \textit{Alice} (\textit{Bob}) randomly selects one of the two disjoint phase intervals $[0,\pi)$ or $[\pi,2\pi)$. Then \textit{she} (\textit{he}) at random chooses a phase from the selected interval and assigns it to $\nu_a$ ($\nu_b$). The phase interval of \textit{Alice} is denoted as $\Delta_a$ and that of \textit{Bob} as $\Delta_b$. Note that $\max(\abs{\nu_a - \nu_b})=\pi$.\\
\textit{Step 3 $'$}. \textit{Measurement}\textemdash \textit{Alice} and \textit{Bob} send their coherent states (signal/test and secondary states) to the intermediate node \textit{Charlie}. At the entry of \textit{Charlie}'s side, in both test and signal rounds, the signal/test coherent states of \textit{Alice} and \textit{Bob} are subjected to an interference at coupler \textbf{C}$_{\text{\textbf{s}}}$, whereas the secondary coherent states of \textit{Alice} and \textit{Bob} are subjected to an interference at coupler \textbf{C}$_{\text{\textbf{r}}}$: the output state of \textbf{C}$_{\text{\textbf{s}}}$ is $\ket{\mathrm{e}^{i\nu_a}\sqrt{\mu} - \mathrm{e}^{i\nu_b}\sqrt{\mu}}$; the output state of \textbf{C}$_{\text{\textbf{s}}}$ is $\ket{\mathrm{e}^{i\nu_a}\sqrt{\mu} - \mathrm{e}^{i\nu_b}\sqrt{\mu}}$. In this protocol (see Fig. \ref{setup-rand}), the secondary coherent states of \textit{Alice} and \textit{Bob} are interchanged in the interference process; see \textit{Step 3} and compare Figs. \ref{setup} and \ref{setup-rand} for verification. The interference process taken place in a coupler is described in Appendix \ref{coupler}. The outputs of the couplers (results of the interference processes just mentioned) are then directed towards a beamsplitter \textbf{BS}. For each round, \textit{Charlie} performs a joint (beamsplitter) measurement that is characterized by POVM elements $F^{\varkappa}$ ~\cite{lin-lutkenhaus}. He announces the measurement outcome $\varkappa \in \{+,-,?,d\}$, where "$+$" identifies the presence of a click only at detector $D_+$, "$-$" identifies the presence of a click only at detector $D_-$, "$?$" identifies the absence of a detector click, and "$d$" identifies the presence of clicks at both detectors ~\cite{lin-lutkenhaus}.
\begin{center}
	\textit{Steps 1 $'$\textendash3 $'$ are iterated $N$ times ($N\rightarrow\infty$), an asymptotic regime of operation is considered}.
\end{center}
\textit{Step 4 $'$}. \textit{Public communication and sifting}\textemdash A public disclosure of information over an authenticated public classical channel takes place in this step. \textit{Alice} and \textit{Bob} first announce \textit{their} values of $m_a$, $m_b$\textemdash\textit{they} perform \textit{round-type} sifting. Next, given a common signal round, \textit{Alice} and \textit{Bob} announce \textit{their} phase intervals $\Delta_a$ and $\Delta_b$\textemdash\textit{they} perform \textit{phase-interval} sifting. In case of a common \textit{signal} round ($m_{a(b)} = 0$) and a common \textit{phase interval} ($\Delta_a=\Delta_b$), both Alice and Bob also performs \textit{outcome sifting}\textemdash\textit{they} keep a common signal round ($m_{a}=m_b=0$) and a common phase interval choice \textit{if and only if} a measurement outcome $\varkappa=-$ is registered and consequently announced by \textit{Charlie}. Note that the signal rounds, retained after the outcome sifting, are used to distill secret-key bits.  \\
In case of a common \textit{test} round ($m_{a(b)} = 1$), \textit{Alice} and \textit{Bob} announce their intensities $\mu_{a(b)}$, their phases $\theta_{a(b)}$ and $\omega_{a(b)}$, and their choices of \textit{sending} or \textit{not sending} (\textit{Alice} and \textit{Bob} thus determines the scenario of this round\textemdash \textbf{sns}, \textbf{nn}, or \textbf{ss} scenario). \textit{They} further perform \textit{scenario sifting} (only \textbf{sns} scenarios are kept), as mentioned earlier. All the test rounds, retained after scenario sifting, are involved in the parameter estimation procedure.\\
\textit{Step 5 $'$}. This step is identical to \textit{Step 5}.\\
\textit{Step 6 $'$}. This step is identical to \textit{Step 6}.\\
\textit{Step 7 $'$}. This step is identical to \textit{Step 7}. \\

\section{Security}\label{security}

\subsection{Security proof}
\begin{figure}[!hbt]
	\centering
	\includegraphics[scale=0.35]{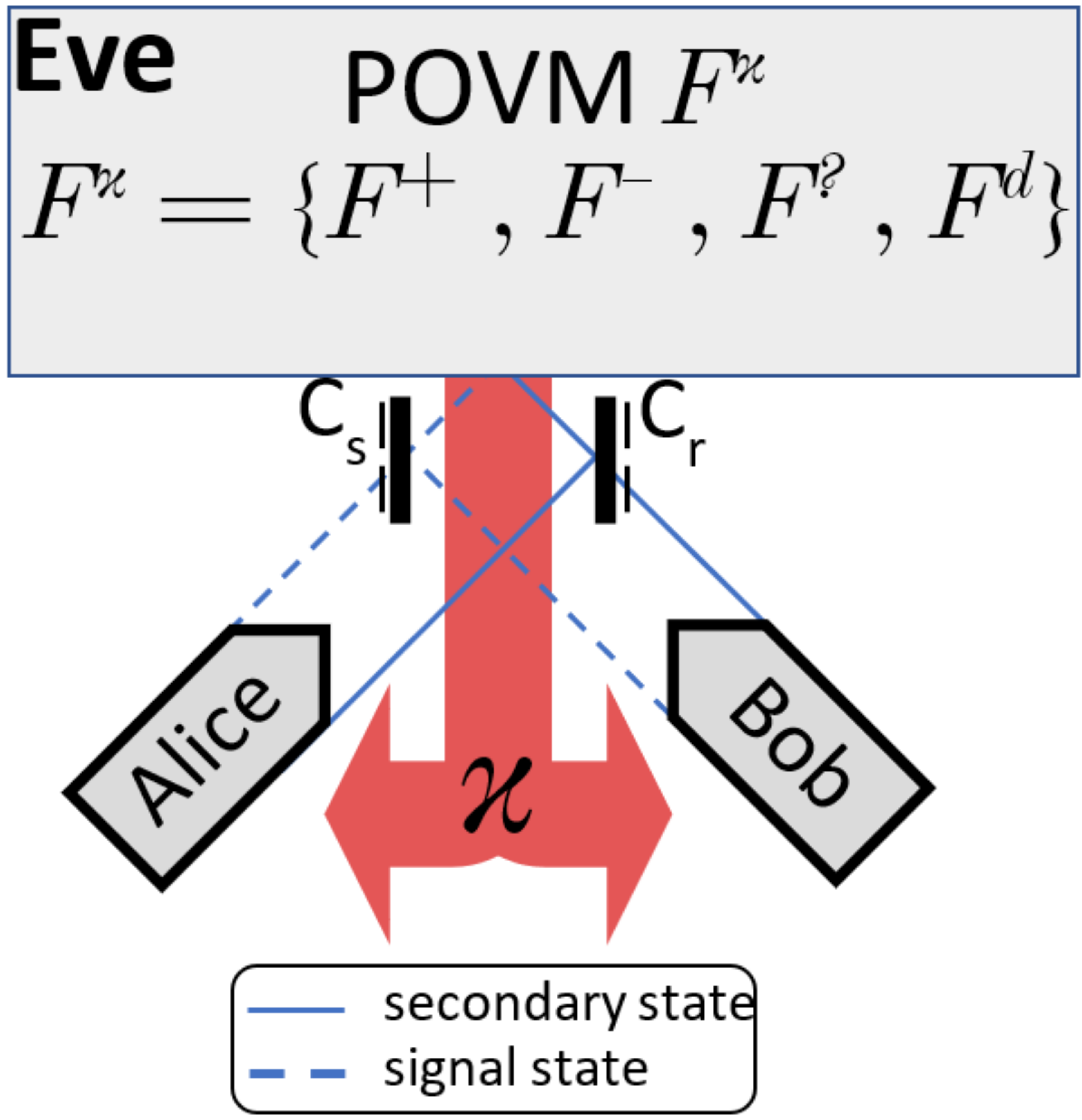}
	\caption{\label{model1} Model of a \textit{POVM attack} on SNS-PM-QKD schematic setup. The attack is in accordance with Ref. ~\cite{lin-lutkenhaus}. \textit{Eve} performs a POVM $F^{\varkappa}$ on the coherent states coming from the couplers' outputs. \textit{Eve} publicly announces the measurement (POVM) outcome $\varkappa\in\{+,-,?,d\}$ ($+$ indicates a click at detector $D_+$ (see Fig. \ref{setup}); $-$ indicates a click at detector $D_-$ (see Fig. \ref{setup}); $?$ indicates the absence of a click at any detector; $d$ indicates the presence of clicks at both detectors). $\delta$---phase mismatch; $\eta_t$---quantum channel transmittance; C$_{\text{x}}$---coupler (x $=$ r---coupler of reference states; x $=$ s---coupler of signal states). }
\end{figure}

\indent This section provides a security proof of the proposed protocols (Sec. \ref{protocol}) against collective attacks given an asymptotic regime of operation (infinite key limit). The security-proof framework of Ref. ~\cite{lin-lutkenhaus} is adopted in this work. The security proof consists of applying the so-called \textit{source-replacement scheme}, applying a \textit{completely positive and trace-preserving map} and performing a \textit{positive operator-valued measurement} (launching the attack), and consequently evaluating the \textit{secret-key generation rate}. \\
\indent In order to gain information about the key bits established by \textit{Alice} and \textit{Bob}, an eavesdropper (Eve) should be aware of  the \textit{sending-or-not-sending} configuration (scenario). We consider a \textit{POVM attack} ~\cite{lin-lutkenhaus}; see Fig. \ref{model1}\textemdash \textit{Eve} observes the outcomes of the joint measurement $F^{\varkappa}$ performed at \textit{Charlie}'s (intermediate node) side.\\
\indent If an eavesdropper (\textit{Eve}) has an access only to the intermediate node of the scheme, the security analysis of the proposed protocol follows the analysis presented in Ref. ~\cite{lin-lutkenhaus}. In this case, \textit{Eve} has an observation only on the states after the couplers (input states of the beamsplitter). Based on these states, \textit{Eve} is not capable of obtaining the key bits established by \textit{Alice} and \textit{Bob}---she gains no knowledge about \textit{sending-or-not-sending} scenario (which
determines the key bit value) of \textit{Alice} and \textit{Bob} when observing only the measurement outcomes $\varkappa$. The security (security proof) of SNS-PM-QKD (as well as SNS-PM-QKD with randomization) is completely identical to that of Ref. ~\cite{lin-lutkenhaus} when \textit{Eve} is restricted only to observing the measurement outcomes.\\
\textit{Note}. The security proof presented below is formulated for SNS-PM-QKD (Sec.~\ref{sns-pm-qkd}); however, it also holds for the phase-randomized variant of the protocol (Sec.~\ref{sns-pm-qkd-r}) provided that both the phase-interval choice and the associated sifting procedure are taken into account. The phase-interval sifting affects only the overall scaling of the secret-key rate obtained from the security proof, \textit{i.e.}, it introduces a multiplicative prefactor $s$, as shown in Eq.~\ref{loss-key-rate-r} (loss-only scenario) and Eq.~\ref{rand-key-rate}.\\
\indent Some preliminaries are put forward in the following lines. In SNS-PM-QKD, \textit{Alice} chooses to send ($w=0$) or not to send ($w=1$) signal $\ket{\varphi}_Q$ and reference $\ket{\tau}_{T}$ coherent states according to \textit{a priori} probability distribution $\{p_w\}$\textemdash the choice is recorded in a classical register $A$ whose states $\{\ket{w}_A\}$ form an \textit{orthonormal} basis. The signal and secondary states of \textit{Alice} are prepared with respect to a round type $u\in\{Z,X\}$ ($Z$$\rightarrow$\textit{signal} round, $X$$\rightarrow$\textit{test} round) according to \textit{a priori} probability distribution $\{p_u\}$\textemdash the choice of a round type is recorded in a classical register $U$ whose states $\{\ket{u}_U\}$ form an \textit{orthonormal} basis. Hence signal $\ket{\varphi}_Q$ and reference $\ket{\tau}_{T}$ coherent states are conditioned on both \textit{sending-or-not-sending} choice and round-type choice, \textit{i.e.}, $\ket{\varphi_{w,u}}_Q$ and $\ket{\tau_{w,u}}_{T}$. \textit{Alice}'s source is then characterized by the following state in the source-replacement scheme ~\cite{ferenczi,lin-lutkenhaus}
\begin{equation}
	\ket{\psi}_{AUQT} = \sum\limits_{w,u}\sqrt{p_w}\sqrt{p_u}\ket{w}_A\ket{u}_U\ket{\varphi_{w,u}}_Q\ket{\tau_{w,u}}_{T}.
\end{equation}
In the proposed protocol, \textit{Bob} chooses to send ($y=0$) or not to send ($y=1$) signal $\ket{\varphi}_{Q'}$ and reference $\ket{\tau}_{T'}$ coherent states according to \textit{a priori} probability distribution $\{q_y\}$\textemdash the choice is recorded in a classical register $B$ whose states $\{\ket{y}_B\}$ form an \textit{orthonormal} basis. The signal and secondary states of \textit{Bob} are prepared with respect to a round type $v\in\{Z,X\}$ ($Z$$\rightarrow$\textit{signal} round, $X$$\rightarrow$\textit{test} round) according to \textit{a priori} probability distribution $\{q_v\}$\textemdash the choice of a round type is recorded in a classical register $V$ whose states $\{\ket{v}_V\}$ form an \textit{orthonormal} basis. Hence signal $\ket{\varphi}_{Q'}$ and reference $\ket{\tau}_{T'}$ coherent states are conditioned on both \textit{sending-or-not-sending} choice and \textit{round-type} choice, \textit{i.e.}, $\ket{\varphi_{y,v}}_{Q'}$ and $\ket{\tau_{y,v}}_{T'}$. \textit{Bob}'s source is then characterized by the following state in the source-replacement scheme ~\cite{ferenczi,lin-lutkenhaus}
\begin{equation}
	\ket{\psi}_{BVQ'T'} = \sum\limits_{y,v}\sqrt{q_y}\sqrt{q_v}\ket{y}_B	\ket{v}_V\ket{\varphi_{y,v}}_T\ket{\tau_{y,v}}_{T'}.
\end{equation}
The overall state of \textit{Alice}'s and \textit{Bob}'s sources is then given by
\begin{equation}
	\begin{split}
		&\ket{\Psi}_{ABUVQTQ'T'} = \ket{\psi}_{AUQT}\otimes\ket{\psi}_{BVQ'T'}	\\
		&= \sum\limits_{w,u,y,v} \sqrt{p_wp_uq_yq_v}\ket{w,y}_{AB}\ket{u,v}_{UV}\ket{\varphi_{w,u},\varphi_{y.v}}_{QQ'}\otimes\\
		&\ket{\tau_{w,u},\tau_{y,v}}_{TT'}
	\end{split}
\end{equation}
Emphasize on that \textit{Eve} has no access to any classical register mentioned above\textemdash neither to \textit{Alice}'s nor to \textit{Bob}'s registers. Note that systems $Q$, $Q'$, $T$, and $T'$ are sent to \textit{Charlie} (\textit{Eve}).
Based on Fig. \ref{model1}, before reaching the beamsplitter (their final destination), system $QQ'$ should pass through coupler \textbf{C$_{\text{s}}$} and system $TT'$ should pass through \textbf{C$_{\text{r}}$}. The \textbf{C$_{\text{s}}$}'s operation, as shown in Appendix \ref{coupler}, is characterized by the following transform
\begin{equation}\label{phi-relation}
	\ket{\varphi_{w,u},\varphi_{y,v}}_{QQ'} \mapsto \ket{\varphi_{w,u}-\mathrm{e}^{i\delta}\sqrt{V}\varphi_{y,v}}_{A'} = \ket{\varphi_{w,y,u,v}}_{A'} 
\end{equation}
$\delta$ being the \textit{phase mismatch} between states of systems $Q$ and $Q'$ at the coupler, and $V$ being the \textit{intensity} (\textit{mode}) \textit{mismatch} between the states. Similarly, the \textbf{C$_{\text{r}}$}'s operation is characterized by the transform
\begin{equation}\label{tau-relation}
	\ket{\tau_{w,u},\tau_{y,v}}_{TT'} \mapsto \ket{\tau_{y,v}-\mathrm{e}^{i\delta}\sqrt{V}\tau_{w,u}}_{B'} = \ket{\tau_{w,y,u,v}}_{B'} 
\end{equation}
Note that $\delta=0$ and $V=1$ in the above expressions when no interference occurs at a coupler (\textit{sending-not-sending} scenario) or ideal (loss-only) scenario is considered. Hence, after couplers \textbf{C$_{\text{s}}$} and \textbf{C$_{\text{r}}$}, the overall state $\ket{\Psi}_{ABUVQTQ'T'}$ becomes
\begin{equation}
	\begin{split}
		&\ket{\Psi}_{ABUVQTQ'T'} \mapsto \ket{\Psi}_{ABUVA'B'}\\
		&= \sum\limits_{w,u,y,v} \sqrt{p_wp_uq_yq_v}\ket{w,y}_{AB}\ket{u,v}_{UV}\ket{\varphi_{w,y,u,v}}_{A'}\otimes\\
		&\ket{\tau_{w,y,u,v}}_{B'} 
	\end{split}
\end{equation} 
The attack of \textit{Eve} is characterized by a CPTP map $\mathcal{E}_{A'B'\rightarrow EC}$ ~\cite{lin-lutkenhaus}
defined as follows
\begin{equation}
	\begin{split}
		\mathcal{E}_{A'B'\rightarrow EC}(\rho) =  \sum\limits_{\varkappa}p_{\varkappa}(\sqrt{F^{\varkappa}}\rho\sqrt{F^{\varkappa}})_E\otimes\ketbra{\varkappa}{\varkappa}_C
	\end{split}
\end{equation}
where $\varkappa$ denotes the measurement outcome being announced, classical register $C$ records the measurement outcome, and $F^{\varkappa}$ are POVM operators used to perform a measurement on system $A'B'$ ~\cite{lin-lutkenhaus} [\textit{Note}: The matrix representations of the POVM operators $F^{\varkappa}$ relevant to this work are provided in Appendix \ref{povm-operators}]. Applied to system $ABUVA'B'$, this map leads to the result
\begin{widetext}
\begin{equation}\label{overall-density}
	\begin{split}
		\rho_{ABUVEC} &= (I_{ABUV}\otimes\mathcal{E}_{A'B'\rightarrow EC})(\ketbra{\Psi}{\Psi}_{ABUVA'B'}) \\
		&= \sum\limits_{w,u,y,v,w',u',y',v'} \sqrt{p_wp_up_{w'}p_{u'}q_yq_vq_{y'}q_{v'}}\ket{w,y}\bra{w',y'}_{AB}\ket{u,v}\bra{u',v'}_{UV}\otimes\\
		&\sum\limits_{\varkappa}(\sqrt{F^{\varkappa}}\ket{\varphi_{w,y,u,v},\tau_{w,y,u,v}}\bra{\varphi_{w',y',u',v'},\tau_{w',y',u',v'}}\sqrt{F^{\varkappa}})_E\otimes\ketbra{\varkappa}{\varkappa}_C
	\end{split}
\end{equation}
\end{widetext}
In accordance with the protocol steps (Sec. \ref{sns-pm-qkd}), \textit{Alice} and \textit{Bob} should perform both \textit{round-type sifting} and \textit{measurement-outcome sifting} before the process of distilling keys (performing a key map) takes place. The round-type sifting consists of announcing variables $u$ and $v$, and discarding those rounds in which $u\neq v$. These actions are characterized by applying an operator of the form
\begin{equation}
	\begin{split}
	\Pi_{\text{round}} &= I_{ABEC}\otimes\ketbra{u=v,v=u}{u=v,v=u}_{UV} \\
	&= I_{ABEC}\otimes\ketbra{u,u}{u,u}_{UV}.
	\end{split}
\end{equation}
Recall that only the case $u=v=Z$ is used for key generation, see the protocol steps of Sec. \ref{sns-pm-qkd}. Therefore
\begin{widetext}
\begin{equation}
	\begin{split}
		\Pi_{\text{round}}(\rho_{ABUVEC})\Pi_{\text{round}} &= \rho_{ABUVEC}' \\
		&= \sum\limits_{w,y,w',y'} p_{u=Z,v=Z}\sqrt{p_wp_{w'}q_yq_{y'}}\ket{w,y}\bra{w',y'}_{AB}\ket{Z,Z}\bra{Z,Z}_{UV}\otimes\\
		&\sum\limits_{\varkappa}(\sqrt{F^{\varkappa}}\ket{\varphi_{w,y,Z,Z},\tau_{w,y,Z,Z}}\bra{\varphi_{w',y',Z,Z},\tau_{w',y',Z,Z}}\sqrt{F^{\varkappa}})_E\otimes\ketbra{\varkappa}{\varkappa}_C
	\end{split}
\end{equation}
\end{widetext}
where $p_{u=Z,v=Z} = p_{u=Z}q_{v=Z}=\sqrt{p_{u=Z}p_{u'=Z}q_{v=Z}q_{v'=Z}}$. Given an asymptotic regime of operation is considered, one can choose $p_{u(v) = Z}\rightarrow1$ (consequently, $p_{u(v) = X}\rightarrow0$). This implies $p_{u=Z,v=Z} \rightarrow 1$ so that 
\begin{widetext}
\begin{equation}
	\begin{split}
		\rho_{ABUVEC}'& \\
		&= \sum\limits_{w,u,y,w',y'} \sqrt{p_wp_{w'}q_yq_{y'}}\ket{w,y}\bra{w',y'}_{AB}\ket{Z,Z}\bra{Z,Z}_{UV}\otimes\\
		&\sum\limits_{\varkappa}(\sqrt{F^{\varkappa}}\ket{\varphi_{w,y,Z,Z},\tau_{w,y,Z,Z}}\bra{\varphi_{w',y',Z,Z},\tau_{w',y',Z,Z}}\sqrt{F^{\varkappa}})_E\otimes\ketbra{\varkappa}{\varkappa}_C
	\end{split}
\end{equation}
\end{widetext}
The measurement outcome sifting consists of announcing variable $\varkappa$ and discarding those rounds in which $\varkappa\neq -$. These actions are characterized by applying an operator of the form
\begin{equation}
	\Pi_{\text{outcome}} = I_{ABUVE}\otimes\ketbra{\varkappa=-}{\varkappa=-}_{C}.
\end{equation}
Recall that only the case $\varkappa=-$ is used for key generation; see the protocol steps of Sec. \ref{protocol}. Hence
\begin{widetext}
\begin{equation}
	\begin{split}
		\Pi_{\text{outcome}}(\rho_{ABUVEC}')\Pi_{\text{outcome}} &= \rho_{ABUVEC}''\\
		&= \sum\limits_{w,y,w',y'} \sqrt{p_wp_{w'}q_yq_{y'}}\ket{w,y}\bra{w',y'}_{AB}\ket{Z,Z}\bra{Z,Z}_{UV}\otimes\\
		&(\sqrt{F^{-}}\ket{\varphi_{w,y,Z,Z},\tau_{w,y,Z,Z}}\bra{\varphi_{w',y',Z,Z},\tau_{w',y',Z,Z}}\sqrt{F^{-}})_E\otimes\ketbra{\varkappa=-}{\varkappa=-}_C
	\end{split}
\end{equation}
\end{widetext}
Now we proceed to the third part of the security proof, namely, evaluation of the secret-key generation rate. Similar to Ref. ~\cite{lin-lutkenhaus}, to evaluate the secret-key rate, we first apply a \textit{key map} to $\rho_{ABUVEC}''$ and then adopt the Devetak-Winter formula ~\cite{devetak}. To establish key bits, \textit{Alice} and \textit{Bob} perform POVMs $M_A$ and $M_B$, respectively\textemdash \textit{Alice} performs measurement $M_A$ ($M_A=\{\ketbra{w}{w}\}$) on system $A$, while \textit{Bob} performs measurement $M_B$ ($M_B=\{\ketbra{y}{y}\}$) on system $B$. The measurement outcome of $M_A$ ($M_B$) is stored in a register $W$ ($Y$). Then a key map $\mathcal{K}$ is applied on system $W$. The map is trivial and characterized by the following transform
\begin{equation}\label{key-map}
	\mathcal{K}_{A\rightarrow K}(\rho_W) = I_W(\ketbra{w}{w}_W)I_W = \ketbra{k}{k}_K.
\end{equation}
Similar to Ref. ~\cite{lin-lutkenhaus}, one performs a CPTP map $\mathcal{G}$ being a combination of the POVMs $M_A$, $M_B$ and the key map $\mathcal{K}$. The result of applying $\mathcal{G}$ to $\rho_{ABUVEC}''$ is
\begin{equation}
	\begin{split}
		&\rho_{KYUVEC}\\
		&=\sum\limits_{k,y}\times p_{\varkappa=-}p_{k,y|\varkappa=-}\ketbra{k}{k}_K\otimes\ketbra{y}{y}_Y\otimes\\
		&\ketbra{Z,Z}{Z,Z}_{UV}\otimes\rho^{k,y,\varkappa,u,v}_E\otimes\ketbra{\varkappa=-}{\varkappa=-}_C\\
		&=\sum\limits_{k,y}\times p_{\varkappa=-}p_{k,y|\varkappa=-}\ketbra{k}{k}_K\otimes\ketbra{y}{y}_Y\otimes\\
		&\ketbra{Z,Z}{Z,Z}_{UV}\otimes\rho^{k,y,-,Z,Z}_E\otimes\ketbra{\varkappa=-}{\varkappa=-}_C.
	\end{split}
\end{equation}
The product of $p_{\varkappa=-}$ and  $p_{k,y|\varkappa=-}$ gives the joint distribution  $p_{k,y,\varkappa=-} = p_{\varkappa=-}\times p_{k,y|\varkappa=-}$.\\
\indent The Devetak-Winter formula ~\cite{devetak} is used to determine the secret-key rate of a QKD protocol under collective attacks. For SNS-PM-QKD, the Devetak-Winter formula is applied to state $\rho_{KYUVEC}$ shared between \textit{Alice}, \textit{Bob}, and \textit{Eve} ~\cite{lin-lutkenhaus}. The state of system $KYUVEC$ can be rewritten as
\begin{eqnarray}
		&&\rho_{KYUVEC} \nonumber\\ &&=p_{\varkappa=-}\times\rho^{\varkappa,u,v}_{KYE}\otimes\ketbra{Z,Z}{Z,Z}_{UV}\otimes\ketbra{\varkappa=-}{\varkappa=-}_C \nonumber\\
		&&=p_{\varkappa=-}\times\rho^{-,Z,Z}_{KYE}\otimes\ketbra{Z,Z}{Z,Z}_{UV}\otimes\ketbra{\varkappa=-}{\varkappa=-}_C \nonumber\\
\end{eqnarray}
where the conditional state shared by \textit{Alice}, \textit{Bob}, and \textit{Eve}
\begin{equation}
	\begin{split}
		\rho^{-,Z,Z}_{KYE} =\sum\limits_{k,y}p_{k,y|\varkappa=-}\ketbra{k}{k}_K\otimes\ketbra{y}{y}_Y\otimes\rho^{k,y,-,Z,Z}_E
	\end{split}
\end{equation}
is conditioned on the announcements $\varkappa$, $u$, and $v$. According to Devetak-Winter formula ~\cite{devetak}, given announcements $\varkappa$, $u$, and $v$, the number of secret-key bits that can be established out of $\rho^{-,Z,Z}_{KYE}$ is ~\cite{lin-lutkenhaus}
\begin{equation}\label{rate-part}
	\begin{split}
		r(\rho^{-,Z,Z}_{KYE}) &= I(K:Y)_{\rho^{-,Z,Z}_{KYE}} - I(K:E)_{\rho^{-,Z,Z}_{KYE}}\\
		&= 1 - \delta^{-,Z,Z}_{\text{EC}} - \chi(K:E)_{\rho^{-,Z,Z}_{KYE}}
	\end{split}
\end{equation}
where $\delta^{-,Z,Z}_{\text{EC}}$~\footnote{This quantity is determined by experimental observations. It is directly proportional to the correlation (anticorrelation) between variables $k$ and $y$.} is the average amount of information leakage during the error correction procedure of a QKD given announcements $\varkappa=-$, $u=Z$, and $v=Z$. This quantity is evaluated by
\begin{equation}\label{error-corr}
	\delta^{-,Z,Z}_{\text{EC}} = f_{\text{EC}}H(e^{-,Z,Z}) 
\end{equation}
$f_{\text{EC}}$ being the \textit{efficiency} of the error correction procedure, $e^{-,Z,Z}$ being the \textit{error rate} in case of $(\varkappa,u,v)$$=$$(-,Z,Z)$, and $H(\cdot)$ being the so-called \textit{Shannon entropy}.
The quantity $\chi(K:E)_{\rho^{-,Z,Z}_{KYE}}$ is the \textit{Holevo information} defined by
\begin{eqnarray}\label{holevo}
		\chi(K:E)_{\rho^{\varkappa,u,v}_{KYE}} &= S(\rho^{\varkappa,u,v}_E) - \sum\limits_{k}p_{k|\varkappa,u,v}\times S(\rho^{k,\varkappa,u,v}_E) \nonumber\\
		&= S(\rho^{-,Z,Z}_E) - \sum\limits_{k}p_{k|-,Z,Z}\times S(\rho^{k,-,Z,Z}_E) \nonumber \\		
\end{eqnarray}
where $S(\rho)$ denotes the so-called von Neumann (quantum) entropy. Note that ~\cite{lin-lutkenhaus}
\begin{equation}\label{conditional-states-rhoe}
	\begin{split}
		\rho^{k,-,Z,Z}_E &= \sum\limits_{y}p_{y|k,-,Z,Z}\times \rho^{k,y,-,Z,Z}_E,\\
		\rho^{-,Z,Z}_E &= \sum\limits_{k} p_{k|-,Z,Z}\times\rho^{k,-,Z,Z}_E.
	\end{split}
\end{equation}
The state $\rho^{k,y,-,Z,Z}_E$ is expressed as
\begin{equation}\label{theta-state-density}
	\rho^{k,y,-,Z,Z}_E = \ketbra{\Theta^{-,Z,Z}_{k,y}}{\Theta^{-,Z,Z}_{k,y}}
\end{equation}
following equation \eqref{overall-density} ~\cite{lin-lutkenhaus}. The conditional state $\ket{\Theta^{-,Z,Z}_{k,y}}$ of \textit{Eve} is defined as
\begin{equation}
	\ket{\Theta^{-,Z,Z}_{k,y}} = \frac{\sqrt{F^-}\ket{\varphi_{k,y,Z,Z},\tau_{k,y,Z,Z}}}{\sqrt{\bra{\varphi_{k,y,Z,Z},\tau_{k,y,Z,Z}}F^{-}\ket{\varphi_{k,y,Z,Z},\tau_{k,y,Z,Z}}}}
\end{equation}
The \textit{total} amount of secret-key bits that \textit{Alice} and \textit{Bob} can establish out of $\rho_{KYUVEC}$ is given by ~\cite{lin-lutkenhaus}
\begin{equation}\label{overall-rate}
	\begin{split}
		R(\rho_{KYUVEC}) &= \sum\limits_{\varkappa,u,v}p_{\varkappa,u,v}\times r(\rho^{\varkappa,u,v}_{KYE}) \\
		&= p_{\varkappa=-,Z,Z}\times r(\rho^{-,Z,Z}_{KYE}).
	\end{split}
\end{equation}

\subsection{Security concerns}
\begin{figure}[!hbt]
	\centering
	\includegraphics[scale=0.25]{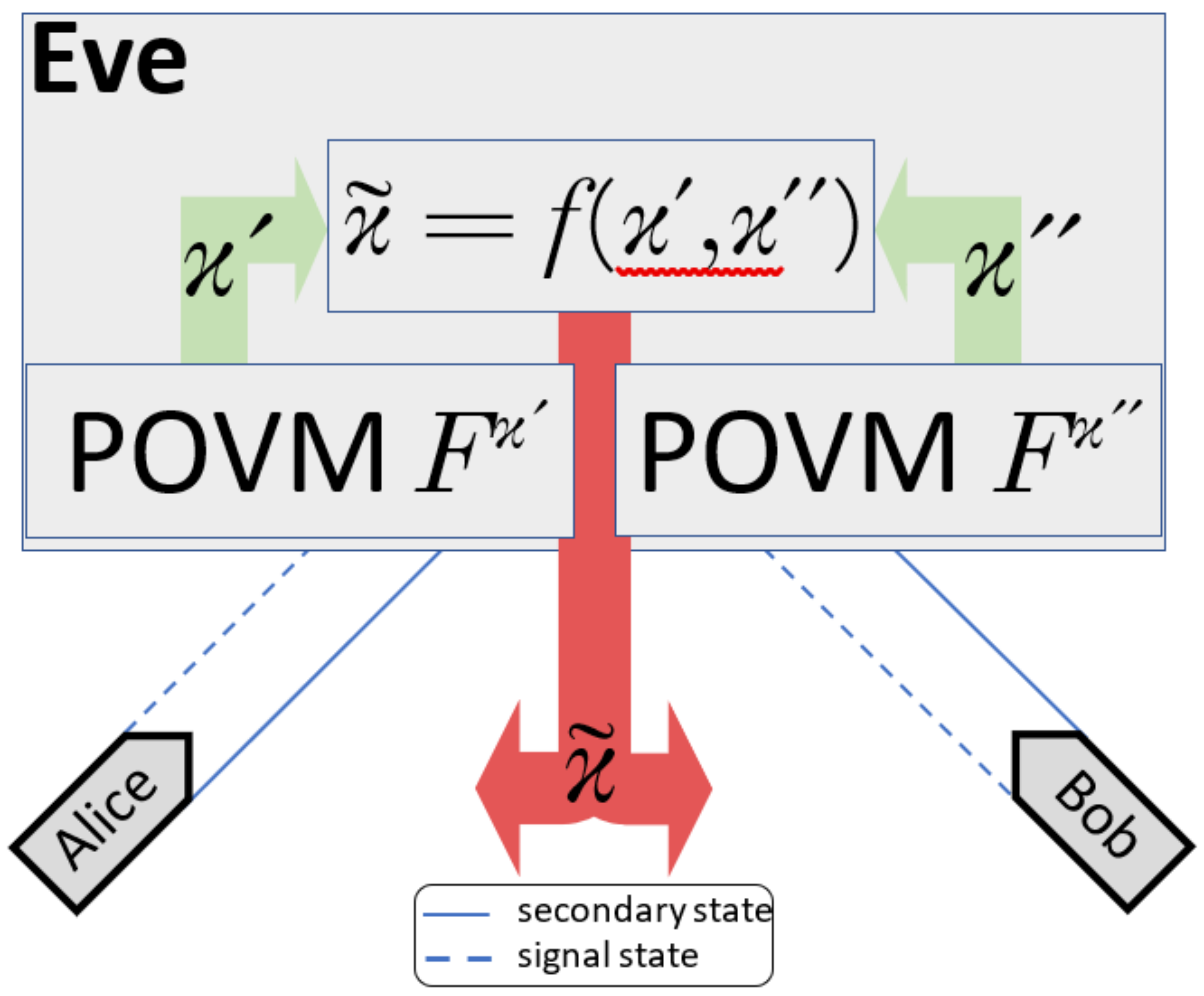}
	\caption{\label{double-povm} Model of a \textit{double-POVM attack} on SNS-PM-QKD schematic setup. \textit{Eve} performs a POVM $F^{\varkappa'}$ on \textit{Alice}'s coherent states and a POVM $F^{\varkappa''}$ on \textit{Bob}'s coherent states. Based on outcomes $\varkappa'$ and $\varkappa''$ of these POVMs, \textit{Eve} decides on the value of $\tilde{\varkappa}$, the actual outcome announcement. \textit{Eve} aims to replicate the outcome announcement $\varkappa$ of the POVM attack (Fig. \ref{model1}).  }
\end{figure}

\begin{figure}[!hbt]
	\centering
	\includegraphics[scale=0.75]{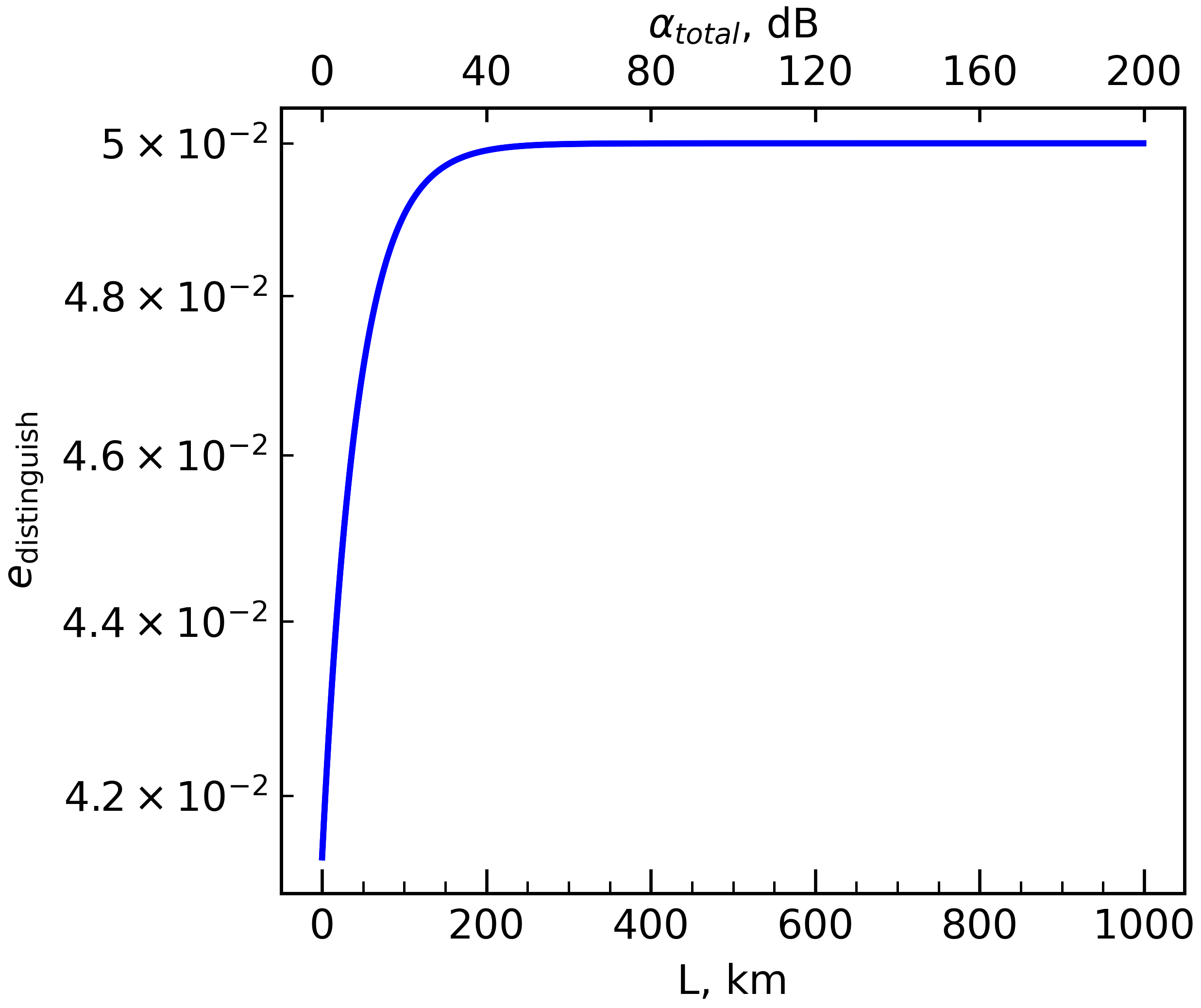}
	\caption{\label{error-double-lo} Error rate $e^{\text{loss-only}}_{\text{distinguish}}$ in a double-POVM attack given a loss-only operation of SNS-PM-QKD. The above curve is evaluated for protocol parameters: channel attenuation $\alpha=0.2$dB, detection efficiency $\eta_{\text{det}}=1$, dark count rate $p_{\text{dark}}=10^{-11}$, mode (intensity) mismatch $V=0.95$, phase mismatch $\delta=\frac{\pi}{60}$, coherent-state intensity  $\mu=0.1$, and sending probability function $\epsilon(L)$ of the form shown in Fig. \ref{epsilon-profile} with $\epsilon_0=0.05$, $\epsilon_{\text{max}}=0.45$, $L_{\text{max}}=950$km. }
\end{figure}

\begin{figure}[!hbt]
	\centering
	\includegraphics[scale=0.75]{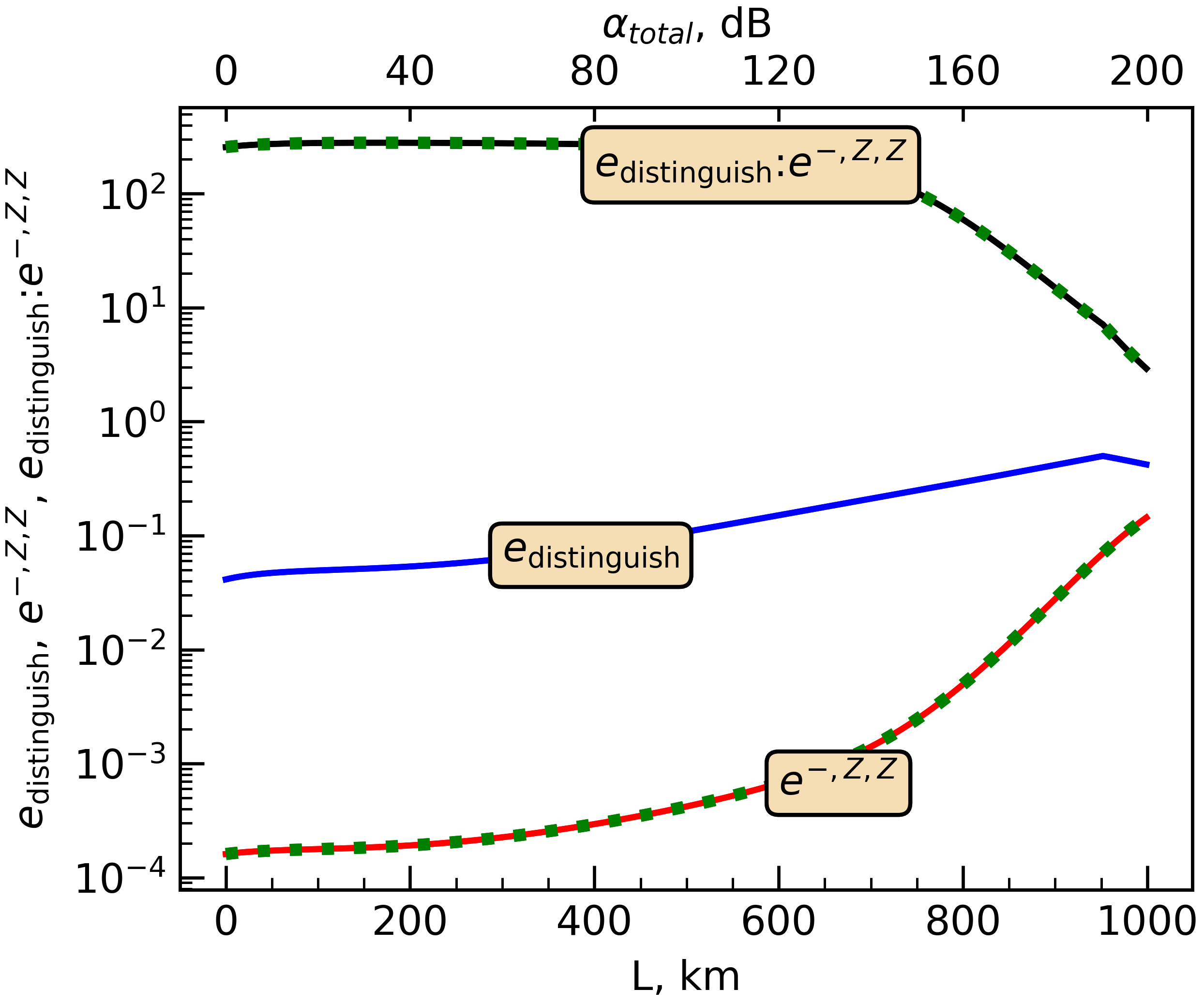}
	\caption{\label{error-double} Comparison between the error rate $e^{-,Z,Z}$ in a POVM attack and the error rate $e^{\text{realistic}}_{\text{distinguish}}$ in a double-POVM attack given a realistic operation of SNS-PM-QKD (SNS-PM-QKD with randomization). The gap between these error rates is quantitatively expressed by the ratio $e^{\text{realistic}}_{\text{distinguish}}:e^{-,Z,Z}$. The error rates and their ratio are evaluated for both SNS-PM-QKD (solid lines) and SNS-PM-QKD with randomization (dotted green lines). Note that $e^{\text{realistic}}_{\text{distinguish}}$ takes identical values for both of these protocols. The above curves are evaluated for protocol parameters: channel attenuation $\alpha=0.2$dB, detection efficiency $\eta_{\text{det}}=1$, dark count rate $p_{\text{dark}}=10^{-11}$, mode (intensity) mismatch $V=0.95$, phase mismatch $\delta=\frac{\pi}{60}$, error correction efficiency $f_{\text{EC}}=1.1$, coherent-state intensity  $\mu=0.1$, and sending probability function $\epsilon(L)$ of the form shown in Fig. \ref{epsilon-profile} with $\epsilon_0=0.05$, $\epsilon_{\text{max}}=0.45$, $L_{\text{max}}=950$km. Note that the relations between the parameters $e^{-,Z,Z}$, $e^{\text{realistic}}_{\text{distinguish}}$, $e^{\text{realistic}}_{\text{distinguish}}:e^{-,Z,Z}$ remain for different values of the protocol parameters as well (e.g., $\alpha=0.2$dB, $\eta_{\text{det}}=0.145$, $p_{\text{dark}}=8\times10^{-8}$, $V=0.95$, $\delta=\frac{\pi}{8}$, $f_{\text{EC}}=1.15$, $\mu=0.1$, $\epsilon(L|\epsilon_0=0.05,\epsilon_{\text{max}}=0.45,L_{\text{max}}=450)$).  }
\end{figure}

\indent This section analyzes an attack specific to the proposed protocol. It is called \textit{double-POVM attack}. Performing this attack, \textit{Eve} aims to learn the \textit{sending-or-not-sending} choices of both \textit{Alice} and \textit{Bob}. The double-POVM attack is illustrated by the scheme of Fig. \ref{double-povm}. As shown in the figure, in this attack, \textit{Eve} performs a POVM $F^{\varkappa'}$ on the coherent states of \textit{Alice} and a POVM $F^{\varkappa''}$ on the coherent states of \textit{Bob}. Based on the outcomes $\varkappa'$ and $\varkappa''$, \textit{Eve} makes a decision about the announcement $\tilde{\varkappa}$\textemdash $\tilde{\varkappa} = f(\varkappa',\varkappa'')$. Note that, in the double-POVM attack, \textit{Eve} attempts to replicate the announcement part of the POVM attack (Fig. \ref{model1}) as closely as possible. \\
\indent The success of the attack lies in distinguishing between scenarios \textbf{sns} and \textbf{ss} (or \textbf{nn}) when the observation $[\varkappa'=+$,$\varkappa''=?]$ or $[\varkappa'=?$,$\varkappa''=+]$ takes place. In other words, the attack is characterized by the error probability of distinguishing \textbf{sns} and \textbf{ss} (or \textbf{nn}), $e_{\text{distinguish}}$\textemdash $e_{\text{distinguish}}$ is actually the error rate induced by \textit{Eve} when double-POVM attack is performed. In what follows, we determine the error probability in order to verify whether or not the double-POVM attack is detectable by \textit{Alice} and \textit{Bob}. Take into account that \textbf{sns} leads to \textit{Alice} and \textit{Bob} sharing correlated key bits, while \textbf{ss} (or \textbf{nn}) leads to \textit{Alice} and \textit{Bob} sharing anticorrelated key bits\textemdash \textbf{ss} (or \textbf{nn}) induces an error. We examine $e_{\text{distinguish}}$ under two scenarios: \textit{loss-only} and \textit{realistic} scenarios. \\
\textit{Loss-only scenario}. To find $e^{\text{loss-only}}_{\text{distinguish}}$, we examine the probability of obtaining the observation $[\varkappa'=+$,$\varkappa''= ?]$~\footnote{Note that the observation $[\varkappa'=?$,$\varkappa''= +]$ leads to identical results.} given \textbf{sns} on one hand, and given \textbf{ss} on the other hand. Take into account that, when loss-only scenario is considered, it is impossible for observation $[\varkappa'=+$,$\varkappa''= ?]$ to take place in case of \textbf{nn} configuration.  We assume that a common signal round is shared by \textit{Alice} and \textit{Bob}. Given $[\varkappa'=+$,$\varkappa''= ?]$, the \textbf{sns} scenario is characterized by \textit{Alice} sending and \textit{Bob} not sending coherent states to the intermediate node. Conditioned on the \textbf{sns} scenario, the probability of obtaining the observation of concern is evaluated by the product
\begin{eqnarray}
		&&p^{\text{\textbf{sns(loss)}}}_{\varkappa'=+,\varkappa''=?}=
		[p_{\text{sending}}\times p_{\varkappa'=+|\text{sending}}]\times\nonumber\\ &&[p_{\text{not-sending}}\times p_{\varkappa''=?|\text{not-sending}}]\\
		&&=[\epsilon\times\bra{+\sqrt{\mu},+\sqrt{\mu}}F^{+'}\ket{+\sqrt{\mu},+\sqrt{\mu}}]\times[(1-\epsilon)\times 1]\nonumber\\
		&&=[\epsilon\times\bra{+\sqrt{\mu},+\sqrt{\mu}}F^{+'}\ket{+\sqrt{\mu},+\sqrt{\mu}}]\times(1-\epsilon)\nonumber
\end{eqnarray}
Notice that $p_{\varkappa'=?|\text{not-sending}}=1$ in the loss-only scenario. The POVM operator $F^{+'}$ is presented in Appendix \ref{povm-operators}. Recall that, in a signal round, \textit{Alice}'s coherent states have the form $\ket{\mathrm{e}^{i\varphi_a}\sqrt{\mu},\mathrm{e}^{i\tau_a}\sqrt{\mu}}_{QT} = \ket{+\sqrt{\mu},+\sqrt{\mu}}$. \\
Given $[\varkappa'=+$,$\varkappa''=?]$, the \textbf{ss} scenario is characterized by both \textit{Alice} and \textit{Bob} sending coherent states to the intermediate node. Conditioned on the \textbf{ss} scenario, the probability of obtaining the observation of concern is evaluated by the product
\begin{equation}
	\begin{split}
		&p^{\text{\textbf{ss(loss)}}}_{\varkappa'=+,\varkappa''=?}\\
		&= [p_{\text{sending}}\times p_{\varkappa'=+|\text{sending}}]\times [p_{\text{sending}}\times p_{\varkappa''=?|\text{sending}}]\\
		&=[\epsilon\times\bra{+\sqrt{\mu},+\sqrt{\mu}}F^{+'}\ket{+\sqrt{\mu},+\sqrt{\mu}}]\times\\
		&[\epsilon\times\bra{+\sqrt{\mu},+\sqrt{\mu}}F^{?''}\ket{+\sqrt{\mu},+\sqrt{\mu}}]
	\end{split}
\end{equation}
The POVM operator $F^{?''}$ is presented in Appendix \ref{povm-operators}. Recall that, in a signal round, \textit{Bob}'s coherent states have the form $\ket{\mathrm{e}^{i\varphi_b}\sqrt{\mu},\mathrm{e}^{i\tau_b}\sqrt{\mu}}_{Q'T'} = \ket{+\sqrt{\mu},+\sqrt{\mu}}$\textemdash they are identical to \textit{Alice}'s.  \\
\indent Hence the error probability of distinguishing \textbf{sns} and \textbf{ss} can be evaluated by
\begin{equation}
	e^{\text{loss-only}}_{\text{distinguish}} = \min\{\tilde{p}^{\text{\textbf{sns(loss)}}}_{\varkappa'=+,\varkappa''=?},\tilde{p}^{\text{\textbf{ss(loss)}}}_{\varkappa'=+,\varkappa''=?}\}
\end{equation}
where 
\begin{equation}
	\begin{split}
		\tilde{p}^{\text{\textbf{sns(loss)}}}_{\varkappa'=+,\varkappa''=?} &= \frac{p^{\text{\textbf{sns(loss)}}}_{\varkappa'=+,\varkappa''=?}}{p^{\text{\textbf{sns(loss)}}}_{\varkappa'=+,\varkappa''=?}+p^{\text{\textbf{ss(loss)}}}_{\varkappa'=+,\varkappa''=?}}, \\
		\tilde{p}^{\text{\textbf{ss(loss)}}}_{\varkappa'=+,\varkappa''=?} &= \frac{p^{\text{\textbf{ss(loss)}}}_{\varkappa'=+,\varkappa''=?}}{p^{\text{\textbf{sns(loss)}}}_{\varkappa'=+,\varkappa''=?}+p^{\text{\textbf{ss(loss)}}}_{\varkappa'=+,\varkappa''=?}}
	\end{split}
\end{equation}
We plot $e^{\text{loss-only}}_{\text{distinguish}}$ as a function of the distance $L$ in Fig. \ref{error-double-lo}. We consider that $\epsilon$ has a constant value ($\epsilon=0.05$). The plot shows that \textit{Eve} induces errors in the raw keys of \textit{Alice} and \textit{Bob} when double-POVM is conducted. Therefore, in order to detect the presence of a double-POVM attack, \textit{Alice} and \textit{Bob} need to sacrifice part of \textit{their} raw-key bits for \textit{parameter estimation} procedure\textemdash \textit{Alice} and \textit{Bob} evaluate the error rate of the quantum channel. If non-zero error rate is observed, \textit{they} proceed to the next step of the proposed protocol (loss-only scenario). Otherwise, \textit{they} terminate the current protocol session and start over a new one.\\
\textit{Realistic scenario}. To find $e^{\text{realistic}}_{\text{distinguish}}$, we examine the probability of obtaining the observation $[\varkappa'=+$,$\varkappa''= ?]$ given \textbf{sns} on one hand, and given \textbf{ss} (or \textbf{nn}) on the other hand. We assume that a common signal round is shared by \textit{Alice} and \textit{Bob}. Given $[\varkappa'=+$,$\varkappa''= ?]$, the \textbf{sns} scenario is characterized by \textit{Alice} sending and \textit{Bob} not sending coherent states to the intermediate node. Conditioned on the \textbf{sns} scenario, the probability of obtaining the observation of concern is evaluated by the product
\begin{equation}
	\begin{split}
		&p^{\text{\textbf{sns(real)}}}_{\varkappa'=+,\varkappa''=?}=
		[p_{\text{sending}}\times p_{\varkappa'=+|\text{sending}}]\times\\ &[p_{\text{not-sending}}\times p_{\varkappa''=?|\text{not-sending}}]\\
		&=[\epsilon\times\bra{+\sqrt{\mu},+\sqrt{\mu}}F^{+'}_{\text{imp}}\ket{+\sqrt{\mu},+\sqrt{\mu}}]\times\\
		&[(1-\epsilon)(1-p_{\text{dark}})^2]
	\end{split}
\end{equation}
The POVM operator $F^{+'}_{\text{imp}}$ is presented in Appendix \ref{povm-operators}. The term $(1-p_{\text{dark}})^2$ signifies that no detector clicks in a POVM $F^{\varkappa''}$ given \textit{Bob} does not send coherent states. Recall that, in a signal round, \textit{Alice}'s coherent states have the form $\ket{\mathrm{e}^{i\varphi_a}\sqrt{\mu},\mathrm{e}^{i\tau_a}\sqrt{\mu}}_{QT} = \ket{+\sqrt{\mu},+\sqrt{\mu}}$.\\
Given $[\varkappa'=+$,$\varkappa''=?]$, the \textbf{ss} scenario is characterized by both \textit{Alice} and \textit{Bob} sending coherent states to the intermediate node. On the other hand the \textbf{nn} scenario is characterized by both \textit{Alice} and \textit{Bob} not sending coherent states to the intermediate node. Conditioned on the \textbf{ss} scenario, the probability of obtaining the observation of concern is evaluated by the product
\begin{equation}
	\begin{split}
		&p^{\text{\textbf{ss(real)}}}_{\varkappa'=+,\varkappa''=?}\\
		&=
		[p_{\text{sending}}\times p_{\varkappa'=+|\text{sending}}]\times [p_{\text{sending}}\times p_{\varkappa''=?|\text{sending}}]\\
		&=[\epsilon\times\bra{+\sqrt{\mu},+\sqrt{\mu}}F^{+'}_{\text{imp}}\ket{+\sqrt{\mu},+\sqrt{\mu}}]\times\\
		&[\epsilon\times\bra{+\sqrt{\mu},+\sqrt{\mu}}F^{?''}_{\text{imp}}\ket{+\sqrt{\mu},+\sqrt{\mu}}]
	\end{split}
\end{equation}
The POVM operator $F^{?''}_{\text{imp}}$ is presented in Appendix \ref{povm-operators}. Recall that, in a signal round, \textit{Bob}'s coherent states have the form $\ket{\mathrm{e}^{i\varphi_b}\sqrt{\mu},\mathrm{e}^{i\tau_b}\sqrt{\mu}}_{Q'T'} = \ket{+\sqrt{\mu},+\sqrt{\mu}}$\textemdash they are identical to \textit{Alice}'s.
Conditioned on the \textbf{nn} scenario, the probability of obtaining the observation of concern is evaluated by the product
\begin{eqnarray}
		&&p^{\text{\textbf{nn(real)}}}_{\varkappa'=+,\varkappa''=?}=
		[p_{\text{not-sending}}\times p_{\varkappa'=+|\text{not-sending}}]\times\nonumber\\ &&[p_{\text{not-sending}}\times p_{\varkappa''=?|\text{not-sending}}]\\
		&&=[(1-\epsilon)\times(1-p_{\text{dark}})p_{\text{dark}}]\times[(1-\epsilon)\times(1-p_{\text{dark}})^2]. \nonumber
\end{eqnarray}
Conditioned on the case "\textbf{ss} (or \textbf{nn})", the probability of obtaining the observation of concern is then evaluated by
\begin{equation}
	\begin{split}
		p^{\text{\textbf{ss$+$nn(real)}}}_{\varkappa'=+,\varkappa''=?}&= p^{\text{\textbf{ss(real)}}}_{\varkappa'=+,\varkappa''=?} + p^{\text{\textbf{nn(real)}}}_{\varkappa'=+,\varkappa''=?}
	\end{split}
\end{equation}
\indent Hence the error probability of distinguishing \textbf{sns} and \textbf{ss} (or \textbf{nn}) can be evaluated by
\begin{equation}
	e^{\text{realistic}}_{\text{distinguish}} = \min\{\tilde{p}^{\text{\textbf{sns(real)}}}_{\varkappa'=+,\varkappa''=?},\tilde{p}^{\text{\textbf{ss$+$nn(real)}}}_{\varkappa'=+,\varkappa''=?}\}
\end{equation}
where 
\begin{equation}
	\begin{split}
		\tilde{p}^{\text{\textbf{sns(real)}}}_{\varkappa'=+,\varkappa''=?} &= \frac{p^{\text{\textbf{sns(real)}}}_{\varkappa'=+,\varkappa''=?}}{p^{\text{\textbf{sns(real)}}}_{\varkappa'=+,\varkappa''=?}+p^{\text{\textbf{ss$+$nn(real)}}}_{\varkappa'=+,\varkappa''=?}}, \\
		\tilde{p}^{\text{\textbf{ss$+$nn(real)}}}_{\varkappa'=+,\varkappa''=?} &= \frac{p^{\text{\textbf{ss$+$nn(real)}}}_{\varkappa'=+,\varkappa''=?}}{p^{\text{\textbf{sns(real)}}}_{\varkappa'=+,\varkappa''=?}+p^{\text{\textbf{ss$+$nn(real)}}}_{\varkappa'=+,\varkappa''=?}}
	\end{split}
\end{equation}
Now we examine the ratio between the error rate of the double-POVM attack, namely $e^{\text{realistic}}_{\text{distinguish}}$, and the error rate $e^{-,Z,Z}$ of the POVM attack (Fig.~\ref{model1}). In doing so, one quantitatively verifies whether \textit{Alice} and \textit{Bob} can register the presence of the double-POVM attack.\\
\indent It is straightforward that $e^{\text{realistic}}_{\text{distinguish}}$ and $e^{-,Z,Z}$ depend only on $L$ provided that the protocol parameters $V$, $\delta$, $\eta_{\text{det}}$, $p_{\text{dark}}$ are fixed (see Eqs.~\eqref{fMinus}-\eqref{abcdop} for reference) and $\epsilon$ is either fixed or regarded as a function of $L$. In Fig.~\ref{error-double}, we compare $e_{\text{distinguish}}$ and $e^{-,Z,Z}$\textemdash the error probabilities are presented as functions of $L$ in the figures. The ratio $e^{\text{realistic}}_{\text{distinguish}}:e^{-,Z,Z}$ is also plotted in this figure. The figure shows that a double-POVM attack leads to an error rate ($e_{\text{distinguish}}$) readily greater than the error rate $e^{-,Z,Z}$ of a POVM attack (Fig.~\ref{model1}) in both SNS-PM-QKD and SNS-PM-QKD with randomization. Hence one can infer that \textit{Alice} and \textit{Bob} are able to register the presence of a double-POVM attack by observing the value of the error rate. \\ 

\section{Numerical simulation}\label{sim}

\subsection{Loss-only scenario}\label{loss-scenario}
\indent In this section, we determine the secret-key rate of SNS-PM-QKD (as well as SNS-PM-QKD with randomization) according to the security proof of Sec.~\ref{security} provided that the quantum channels \textit{Alice}-\textit{Charlie} and \textit{Bob}-\textit{Charlie} are regarded as lossy, noiseless channels (\textit{loss-only scenario}). We assume that \textit{Alice} and \textit{Bob} are located at equal distances from \textit{Charlie}. The loss-only scenario implies that efficiencies of detectors are $\eta_{\text{det}}=1$, dark count rates of detectors are $p_d = 0$, channel transmittance $\eta_{\text{t}}=10^{-\frac{\alpha L}{10}}=10^{-\frac{0.2 L}{10}}$ (\textit{i.e.}, channel attenuation is $\alpha=0.2$), beamsplitter and couplers operate in an ideal manner\textemdash no phase mismatch is present ($\delta = 0$). Hence, based on the descriptions of the proposed protocols (see Sec.~\ref{protocol}), only \textbf{sns} configurations lead to relevant measurement outcomes  (announcements) $\varkappa$ ($\varkappa=-$) in the loss-only scenario. Note that \textbf{sns} configurations appear with probability $2\epsilon(1-\epsilon)$ in overall\textemdash $\epsilon(1-\epsilon)$ for \textit{sending-not sending} and $(1-\epsilon)\epsilon$ for \textit{not sending-sending} configuration. This implies that any relevant outcome in the loss-only scenario occurs with probability $2\epsilon(1-\epsilon)\times p_{\varkappa=-,Z,Z|\varphi_{a(b)},\tau_{a(b)}}$ where $\varphi_{a(b)}$ and $\tau_{a(b)}$ denote the signal and secondary coherent states chosen by \textit{Alice} (\textit{Bob}), $u$ ($v$) denotes the basis chosen and announced by \textit{Alice} (\textit{Bob}). Since loss-only scenario implies that there is a lack of noise in SNS-PM-QKD, the amount of information leakage during error correction procedure is $\delta^{\varkappa,u,v}_{EC}=0$. That is, error correction is not performed.\\
\indent To determine the secret-key rate of SNS-PM-QKD, we are required to obtain the Holevo information $\chi(K:E)$ given $\varkappa=-$, $u=Z$, and $v=Z$ (key-generation rounds). For this purpose, $\rho^{k,y,-,Z,Z}_E$, $\rho^{k,-,Z,Z}_E$, and $\rho^{-,Z,Z}_E$ are to be found; see expression \eqref{holevo}. Using expression \eqref{conditional-states-rhoe}, we find that
\begin{equation}\label{rhoe-cond-1}
	\begin{split}
		\rho^{k,-,Z,Z}_E &= \sum\limits_y p_{y|k,-,Z,Z}\times\rho^{k,y,-,Z,Z}_E\\
		&=p_{y=\overline{k}|k,-,Z,Z}\times\rho^{k,y=\overline{k},-,Z,Z}_E\\
		&=1\times\rho^{k,y=\overline{k},-,Z,Z}_E \overset{\eqref{theta-state-density}}{=} \ketbra{\Theta^{-,Z,Z}_{k,y=\overline{k}}}{\Theta^{-,Z,Z}_{k,y=\overline{k}}}
	\end{split}
\end{equation}
and
\begin{eqnarray}
		\rho^{-,Z,Z}_E &&= \sum\limits_k p_{k|-,Z,Z}\times\rho^{k,-,Z,Z}_E\\
		&& = p_{0|-,Z,Z}\times\rho^{0,-,Z,Z}_E + p_{1|-,Z,Z}\times\rho^{1,-,Z,Z}_E \nonumber\\
		&&= \frac{1}{2}\ketbra{\Theta^{-,Z,Z}_{0,1}}{\Theta^{-,Z,Z}_{0,1}} + \frac{1}{2}\ketbra{\Theta^{-,Z,Z}_{1,0}}{\Theta^{-,Z,Z}_{1,0}}. \nonumber
\end{eqnarray}
The result of equation \eqref{rhoe-cond-1} implies that
$\rho^{k,-,Z,Z}_E$ $\forall k\in\{0,1\}$ is a pure state, \textit{i.e.}, its von Neumann entropy $S(\rho^{k,-,Z,Z}_E)$ vanishes\textemdash $S(\rho^{k,-,Z,Z}_E)=0$; in accordance with Ref.~\cite{lin-lutkenhaus}. Notice above that  $p_{y=\overline{k}|k,-,Z,Z}=1$. This follows from the fact that either $k=\overline{y}$ or $y=\overline{k}$ is possible given announcements $\varkappa=-$, $u=Z$, and $v=Z$ in the loss-only scenario. Also notice the identity $p_{0|-,Z,Z}=p_{1|-,Z,Z}$. It follows from the fact that \textit{Alice} and \textit{Bob} send (do not send)  states with identical probability $\epsilon$ ($1-\epsilon$). That is, in a \textbf{sns} configuration, \textit{Alice} sends coherent states in one half of the cases and \textit{Bob} sends coherent states in the other half of the cases. The possible $\rho^{k,y,-,Z,Z}_E$ states are given by
\begin{equation}
	\begin{split}
		\rho^{0,1,-,Z,Z}_E &= \ketbra{\Theta^{-,Z,Z}_{0,1}}{\Theta^{-,Z,Z}_{0,1}}\\
		\rho^{1,0,-,Z,Z}_E &=\ketbra{\Theta^{-,Z,Z}_{1,0}}{\Theta^{-,Z,Z}_{1,0}} 
	\end{split}
\end{equation}
where
\begin{eqnarray}
		\ket{\Theta^{-,Z,Z}_{0,1}} &&= \frac{\sqrt{F^-}\ket{\varphi_{0,1,Z,Z},\tau_{0,1,Z,Z}}}{\sqrt{\bra{\varphi_{0,1,Z,Z},\tau_{0,1,Z,Z}}F^{-}\ket{\varphi_{0,1,Z,Z},\tau_{0,1,Z,Z}}}} \nonumber\\
		&& = \frac{\sqrt{F^-}\ket{+\sqrt{\mu},-\sqrt{\mu}}}{\sqrt{\bra{+\sqrt{\mu},-\sqrt{\mu}}F^{-}\ket{+\sqrt{\mu},-\sqrt{\mu}}}}, \nonumber\\
		\ket{\Theta^{-,Z,Z}_{1,0}} &&= \frac{\sqrt{F^-}\ket{\varphi_{1,0,Z,Z},\tau_{1,0,Z,Z}}}{\sqrt{\bra{\varphi_{1,0,Z,Z},\tau_{1,0,Z,Z}}F^{-}\ket{\varphi_{1,0,Z,Z},\tau_{1,0,Z,Z}}}} \nonumber\\
		&& = \frac{\sqrt{F^-}\ket{-\sqrt{\mu},+\sqrt{\mu}}}{\sqrt{\bra{-\sqrt{\mu},+\sqrt{\mu}}F^{-}\ket{-\sqrt{\mu},+\sqrt{\mu}}}}. \nonumber\\
\end{eqnarray}
Given announcements $\varkappa=-$, $u=Z$, and $v=Z$, the coherent states $\ket{\varphi_{k,y,Z,Z},\tau_{k,y,Z,Z}}$ take the following forms according to the key map \eqref{key-map} and relations  \eqref{phi-relation}-\eqref{tau-relation}
\begin{equation}
	\begin{split}
		&\ket{\varphi_{k,y,Z,Z},\tau_{k,y,Z,Z}} \\
		&= \ket{\varphi_{w,y,Z,Z},\tau_{w,y,Z,Z}} = 
		\ket{\varphi_{0,1,Z,Z},\tau_{0,1,Z,Z}} \\
		&= \ket{\varphi_{k,Z}-\mathrm{e}^{i\delta}\sqrt{V}\varphi_{y,Z},\tau_{y,Z}-\mathrm{e}^{i\delta}\sqrt{V}\tau_{k,Z}}\\
		&=  \ket{\varphi_{0,Z}-\mathrm{e}^{i\delta}\sqrt{V}\varphi_{1,Z},\tau_{1,Z}-\mathrm{e}^{i\delta}\sqrt{V}\tau_{0,Z}}\\
		&\overset{\delta=0,V=1}{=}\ket{\varphi_{0,Z},-\tau_{0,Z}} = \ket{\sqrt{\mu},-\sqrt{\mu}},\\
		&\ket{\varphi_{k,y,Z,Z},\tau_{k,y,Z,Z}} \\
		&= \ket{\varphi_{w,y,Z,Z},\tau_{w,y,Z,Z}} = \ket{\varphi_{1,0,Z,Z},\tau_{1,0,Z,Z}} \\
		&= \ket{\varphi_{1,Z}-\mathrm{e}^{i\delta}\sqrt{V}\varphi_{0,Z},\tau_{0,Z}-\mathrm{e}^{i\delta}\sqrt{V}\tau_{1,Z}}\\
		&\overset{\delta=0,V=1}{=}\ket{-\varphi_{0,Z},\tau_{0,Z}} = \ket{-\sqrt{\mu},\sqrt{\mu}}.
	\end{split}
\end{equation}
In the above expressions, $\delta=0$ and $V=1$ when ideal (loss-only) scenario is considered, as aforementioned in the text.
Recall that the \textbf{sns} configuration ($k=\overline{y}$ or $y=\overline{k}$) is inferred from announcement $\varkappa=-$ itself. It is straightforward that the von Neumann entropy $S(\rho^{-,Z,Z}_E)$ is equal to $H\left(\frac{1-	\abs{\braket{\Theta^{-,Z,Z}_{0,1}}{\Theta^{-,Z,Z}_{1,0}}}}{2}\right)$ where $H(\cdot)$ is the Shannon entropy ~\cite{lin-lutkenhaus}. Take into account that the eigenvalues of $\rho^{-,Z,Z}_E$ are $\frac{1}{2}\left(1\pm\abs{\braket{\Theta^{-,Z,Z}_{0,1}}{\Theta^{-,Z,Z}_{1,0}}}\right)$. According to Ref. ~\cite{lin-lutkenhaus}, 
\begin{equation}\label{entropy-rhoe}
	\begin{split}
		H\left(\frac{1-	\abs{\braket{\Theta^{-,Z,Z}_{0,1}}{\Theta^{-,Z,Z}_{1,0}}}}{2}\right) \\
		= H\left(\frac{1-\mathrm{e}^{-4\mu(1-\sqrt{\eta})}\mathrm{e}^{-2\mu\sqrt{\eta}}}{2}\right).
	\end{split}
\end{equation}
\indent We are now in a position to evaluate $\chi(K:E)$ by using definition \eqref{holevo}.
Taking into consideration result \eqref{entropy-rhoe} and $S(\rho^{k,-,Z,Z}_E)=0$, we obtain for $\chi(K:E)$
\begin{equation}
	\chi(K:E) = H\left(\frac{1-\mathrm{e}^{-4\mu(1-\sqrt{\eta})}\mathrm{e}^{-2\mu\sqrt{\eta}}}{2}\right).
\end{equation}
Hence, according to definition \eqref{rate-part}, the rate $r(\rho^{-,Z,Z}_{KYE})$ takes the form
\begin{equation}
	\begin{split}
		r(\rho^{-,Z,Z}_{KYE}) &= 1 - \delta^{-,Z,Z}_{\text{EC}} - \chi(K:E)\\
		&= 1  - H\left(\frac{1-\mathrm{e}^{-4\mu(1-\sqrt{\eta})}\mathrm{e}^{-2\mu\sqrt{\eta}}}{2}\right).
	\end{split}
\end{equation}
According to \eqref{overall-rate}, the overall rate $R$ is then
\begin{equation}\label{loss-key-rate}
	\begin{split}
		R &= p_{-,Z,Z}\times r(\rho^{-,Z,Z}_{KYE})
		= 2\epsilon(1-\epsilon)(1-\mathrm{e}^{-2\mu\sqrt{\eta}})\times\\
		&\left[1  - H\left(\frac{1-\mathrm{e}^{-4\mu(1-\sqrt{\eta})}\mathrm{e}^{-2\mu\sqrt{\eta}}}{2}\right)\right].
	\end{split}
\end{equation}
The overall rate of a SNS-PM-QKD with phase randomization is then
\begin{equation}\label{loss-key-rate-r}
	\begin{split}
		R' &= s\times2\epsilon(1-\epsilon)(1-\mathrm{e}^{-2\mu\sqrt{\eta}})\times\\
		&\left[1  - H\left(\frac{1-\mathrm{e}^{-4\mu(1-\sqrt{\eta})}\mathrm{e}^{-2\mu\sqrt{\eta}}}{2}\right)\right].
	\end{split}
\end{equation}
where $s$ ($s=0.5$) is the \textit{phase-interval sifting factor}\textemdash the probability of \textit{Alice} and \textit{Bob} picking the same phase interval ($[0,\pi)$ or $[\pi,2\pi)$). \\
\indent The value of $p_{-,Z,Z}$ above is obtained by taking into account Table \ref{table} applied to the following expression
\begin{eqnarray}\label{sum-prob}
		&&p_{\varkappa,u,v} \nonumber\\
		&&= p_{-,Z,Z} =\sum\limits_{\varphi,\tau} p_{-,Z,Z|\varphi,\tau}\times p_{\varphi,\tau} \\
		&&=p_{-,Z,Z|+\sqrt{\mu},-\sqrt{\mu}}\times p_{+\sqrt{\mu},-\sqrt{\mu}} 
		+p_{-,Z,Z|-\sqrt{\mu},+\sqrt{\mu}}\times \nonumber\\
		&&p_{-\sqrt{\mu},+\sqrt{\mu}}=(1-\mathrm{e}^{-2\mu\sqrt{\eta}})2\epsilon(1-\epsilon). \nonumber
\end{eqnarray}
Probabilities $p_{-,Z,Z|+\sqrt{\mu},-\sqrt{\mu}}$ and $p_{-,Z,Z|-\sqrt{\mu},+\sqrt{\mu}}$ are evaluated by expressions ~\cite{lin-lutkenhaus}
\begin{eqnarray}
		&&p_{-,Z,Z|+\sqrt{\mu},-\sqrt{\mu}} \nonumber\\
		&&=\bra{+\sqrt{\mu},-\sqrt{\mu}}F^-\ket{+\sqrt{\mu},-\sqrt{\mu}} \nonumber \\
		&&= \mathrm{e}^{-\frac{\sqrt{\eta}\abs{+\sqrt{\mu}+(-\sqrt{\mu})}^2}{2}}(1-\mathrm{e}^{-\frac{\sqrt{\eta}\abs{+\sqrt{\mu}-(-\sqrt{\mu})}^2}{2}}) \\
		&&p_{-,Z,Z|-\sqrt{\mu},+\sqrt{\mu}} \nonumber\\
		&&=\bra{-\sqrt{\mu},+\sqrt{\mu}}F^-\ket{-\sqrt{\mu},+\sqrt{\mu}} \nonumber \\ &&= \mathrm{e}^{-\frac{\sqrt{\eta}\abs{-\sqrt{\mu}+(+\sqrt{\mu})}^2}{2}}(1-\mathrm{e}^{-\frac{\sqrt{\eta}\abs{-\sqrt{\mu}-(+\sqrt{\mu})}^2}{2}}) \nonumber 		
\end{eqnarray}
Probabilities $p_{\varphi,\tau}$ are related to the choice of \textit{sending-or-not-sending}:  
\begin{equation}
	\begin{split}
		p_{\varphi,\tau} &= p_{\varphi=\varphi_{0,1,Z,Z},\tau=\tau_{0,1,Z,Z}}=\epsilon(1-\epsilon) \\ &\text{given \textit{sending-not sending}} \;\;\; \text{\textbf{sns}},\\
		p_{\varphi,\tau} &= p_{\varphi=\varphi_{1,0,Z,Z},\tau=\tau_{1,0,Z,Z}}=(1-\epsilon)\epsilon \\ &\text{given \textit{not sending-sending}} \;\;\; \text{\textbf{sns}},\\
		p_{\varphi,\tau} &= p_{\varphi=\varphi_{1,1,Z,Z},\tau=\tau_{1,1,Z,Z}} = (1-\epsilon)^2 \\
		&\text{given \textit{not sending-not sending}} \;\;\; \text{\textbf{nn}},\\
		p_{\varphi,\tau} &= p_{\varphi=\varphi_{0,0,Z,Z},\tau=\tau_{0,0,Z,Z}} = \epsilon^2 \\ &\text{given \textit{sending-sending}} \;\;\; \text{\textbf{ss}}
	\end{split}
\end{equation}
In configurations \textbf{nn} and \textbf{ss}, no clicks are registered at $D_-$ detector when SNS-PM-QKD (or SNS-PM-QKD with randomization) is considered to operate in a loss-only scenario. Therefore $\varkappa=-$ does not occurs in these configurations. For SNS-PM-QKD this follows from the fact that
\begin{equation*}
	\begin{split}
		&\ket{\varphi_{1,1,Z,Z},\tau_{1,1,Z,Z}}\\
		&=\ket{\varphi_{1,Z}-\mathrm{e}^{i\delta}\sqrt{V}\varphi_{1,Z},\tau_{1,Z}-\mathrm{e}^{i\delta}\sqrt{V}\tau_{1,Z}} \\
		&\overset{\delta=0,V=1}{=} \ket{\sqrt{0}-\sqrt{0},\sqrt{0}-\sqrt{0}} = \ket{0,0},\\
		&\ket{\varphi_{0,0,Z,Z},\tau_{0,0,Z,Z}}\\
		&=\ket{\varphi_{0,Z}-\mathrm{e}^{i\delta}\sqrt{V}\varphi_{0,Z},\tau_{0,Z}-\mathrm{e}^{i\delta}\sqrt{V}\tau_{0,Z}} \\
		&\overset{\delta=0,V=1}{=} \ket{+\sqrt{\mu}-(+\sqrt{\mu}),+\sqrt{\mu}-(+\sqrt{\mu})} \\
		&= \ket{0,0}
	\end{split}
\end{equation*}
therefore
\begin{equation*}
	\begin{split}
		p_{-,Z,Z|0,0}^{\text{nn}} &= \bra{0,0}F^-\ket{0,0} = 0,\\
		p_{-,Z,Z|0,0}^{\text{ss}} &= \bra{0,0}F^-\ket{0,0} = 0
	\end{split}
\end{equation*}
As a result, both $p_{-,Z,Z|0,0}^{\text{nn}}p_{\varphi=\varphi_{1,1,Z,Z},\tau=\tau_{1,1,Z,Z}}$ and $p_{-,Z,Z|0,0}^{\text{ss}}p_{\varphi=\varphi_{0,0,Z,Z},\tau=\tau_{0,0,Z,Z}}$ vanish and do not contribute to the sum of Eq. \eqref{sum-prob}. Similarly, for SNS-PM-QKD with phase randomization we have
\begin{equation*}
	\begin{split}
		p_{-,Z,Z|0,0}^{\text{nn}} &= \bra{0,0}F^-\ket{0,0} = 0,\\
		p_{-,Z,Z|\sqrt{\psi},\sqrt{\psi}}^{\text{ss}} &= \bra{\sqrt{\psi},\sqrt{\psi}}F^-\ket{\sqrt{\psi},\sqrt{\psi}} = 0
	\end{split}
\end{equation*}
where $\sqrt{\psi}$ denotes the intensity of a coupler output state for the \textbf{ss} configuration; see Appendix~\ref{coupler} for reference.\\
\indent The rate-distance functions $R(L)$ and $R'(L)$ related to loss-only scenario are depicted in Sec. \ref{results}.

\begin{table}[b]
	\caption{Conditional probability distribution of announcements $\varkappa$, $u=Z$, and $v=Z$  given the possible coherent states (signal and reference states) in the loss-only scenario. Note that only the key-generation scenario ($u=Z$ and $v=Z$) is considered. $\ket{\varphi_{0,1,Z,Z},\tau_{0,1,Z,Z}}$ corresponds to $\ket{+\sqrt{\mu},-\sqrt{\mu}}$ and $\ket{\varphi_{1,0,Z,Z},\tau_{1,0,Z,Z}}$ corresponds to $\ket{-\sqrt{\mu},+\sqrt{\mu}}$ in the table. Take into account that this table completely resembles that of Ref. ~\cite{lin-lutkenhaus}\textemdash coherent states $\ket{\varphi,\tau}$ and POVM $F^{\varkappa}$ of this work are identical to those of ~\cite{lin-lutkenhaus}.}\label{table}
	\centering
	\begin{tabular}{c|c|c}
		\hline
		\hline
		$\ket{\varphi,\tau}$ & $\ket{\varphi_{0,1,Z,Z},\tau_{0,1,Z,Z}}$ & $\ket{\varphi_{1,0,Z,Z},\tau_{1,0,Z,Z}}$ \\
		\hline
		$p_{-,Z,Z|\varphi,\tau}$ & $1-\mathrm{e}^{-2\mu\sqrt{\eta}}$ & $1-\mathrm{e}^{-2\mu\sqrt{\eta}}$ \\
		$p_{?,Z,Z|\varphi,\tau}$ & $\mathrm{e}^{-2\mu\sqrt{\eta}}$ & $\mathrm{e}^{-2\mu\sqrt{\eta}}$ \\
		\hline\hline
	\end{tabular}
\end{table}

\subsection{Realistic scenario}\label{imperfections}
\indent In this section, the behavior of the proposed protocol is examined in case of  realistic scenario (presence of imperfections in the communication setup). Similar to ~\cite{lin-lutkenhaus}, the realistic scenario involves: non-ideal \textit{detection efficiency} $\eta_{\text{det}}$ of detectors ($\eta_{\text{det}}<1$), non-zero \textit{dark count rate} $p_{\text{dark}}$ of detectors ($p_{\text{dark}}\neq0$), non-zero \textit{phase mismatch} $\delta$ in an interference process occurring at a beamsplitter or a coupler ($\delta\neq0$), non-ideal \textit{mode} (\textit{intensity}) \textit{mismatch} $V$ in an interference process ($V<1$),  non-ideal \textit{error correction efficiency} $f_{\text{EC}}$ ($f_{\text{EC}}>1$). Note that these imprefections are used to model the non-ideal behavior of all the parts of the communication setup shown in Fig. \ref{setup}.\\
\indent Two of the quantities above are used to characterize the imperfections of the detection system (detectors)\textemdash these are detection efficiency $\eta_{\text{det}}$ and dark count rate of the detectors $p_{\text{dark}}$.
We assume that both detectors of the setup (see Fig. \ref{setup}) have identical efficiencies ($\eta_{\text{det}}^{D_+}=\eta_{\text{det}}^{D_-}$$\rightarrow \eta_{\text{det}}^2$ in the expression of $\eta$ below) and identical dark count rates ($p_{\text{dark}}^{D_+}=p_{\text{dark}}^{D_-}$).
The detector efficiency is involved in the \textit{overall} transmittance of the setup $\eta=\eta_{\text{det}}^2\eta_{\text{t}}=\eta_{\text{det}}^210^{-\alpha L/10}$ where $\eta_{\text{t}}$ denotes \textit{channel transmittance}, $\alpha$ [in dB/km] denotes \textit{channel attenuation per kilometer}, and $L$ [in km] is the \textit{transmission} (\textit{operational}) \textit{distance} of the setup. \\
\indent Both phase mismatch and mode (intensity) mismatch are involved in any interference process occurring in the setup. These quantities are used to characterize the asymmetry between coherent states in an interference process. The asymmetry is caused by the non-ideal behavior of either a beamsplitter, or a coupler, or the quantum channel. The quantum channel could induce the so-called \textit{phase drift} into a coherent state. At a beamsplitter, phase and intensity mismatches are taken into account according to the security-proof analysis of ~\cite{lin-lutkenhaus}\textemdash the state arriving at a beamsplitter is $\ket{\varphi,\mathrm{e}^{i\delta}\tau}$ when phase mismatch is considered, and the arriving state is $\ket{\varphi,\sqrt{V}\tau}$ in the original mode and $\ket{0,\sqrt{1-V}\tau}$ in the second mode when mode (intensity) mismatch is considered. At a coupler, phase and intensity (mode) mismatches are taken into account as described in Appendix \ref{coupler}.\\
\indent In case of setup imperfections, the overall rate of SNS-PM-QKD (Sec.~\ref{sns-pm-qkd}) becomes
\begin{equation}\label{real-key-rate}
	\begin{split}
		R_{\text{real}} &= p_{-,Z,Z}\times r(\rho^{-,Z,Z}_{KYE}) \\ &= \left(2\epsilon(1-\epsilon)P_{\text{sns}} + \epsilon^2P_{\text{ss}} + (1-\epsilon)^2P_{\text{nn}}\right)\times\\
		&[1 - \chi(K:E)^{-,Z,Z} - \delta^{-,Z,Z}_{\text{EC}}].
	\end{split}
\end{equation}
The quantity $P_{\text{sns}}$ is the probability of $\varkappa=-$ to occur in a $Z$ (signal) round given \textit{Alice} sends and \textit{Bob} does not send coherent states \textit{or} \textit{Alice} does not send and \textit{Bob} sends coherent states. It is given by 
\begin{equation}
	\begin{split}
		P_{\text{sns}} &= \bra{\varphi_{0,1,Z,Z},\tau_{0,1,Z,Z}}F^{-}_{\text{imp}}\ket{\varphi_{0,1,Z,Z},\tau_{0,1,Z,Z}} + \\ &\bra{\varphi_{1,0,Z,Z},\tau_{1,0,Z,Z}}F^{-}_{\text{imp}}\ket{\varphi_{1,0,Z,Z},\tau_{1,0,Z,Z}}\\
		&= \bra{+\sqrt{\mu},-\sqrt{\mu}}F^{-}_{\text{imp}}\ket{+\sqrt{\mu},-\sqrt{\mu}} +\\ &\bra{-\sqrt{\mu},+\sqrt{\mu}}F^{-}_{\text{imp}}\ket{-\sqrt{\mu},+\sqrt{\mu}}.
	\end{split}
\end{equation}
$F^{-}_{\text{imp}}$ is a POVM operator corresponding to outcome $\varkappa=-$ in case of imperfections ~\cite{lin-lutkenhaus}. The matrix form of $F^{-}_{\text{imp}}$ is shown in Appendix \ref{povm-operators}. The quantity $P_{\text{ss}}$ is the probability of $\varkappa=-$ to occur given both \textit{Alice} and \textit{Bob} send coherent states in a $Z$ (signal) round. It is given by
\begin{equation}
	\begin{split}
		P_{\text{ss}} &= \bra{\sqrt{\gamma},\sqrt{\gamma}}F^{-}_{\text{imp}}\ket{\sqrt{\gamma},\sqrt{\gamma}}
	\end{split}
\end{equation}
where $\ket{\sqrt{\gamma}}$ signifies the output state of a coupler in case of phase mismatch  $\delta$ between input states $\ket{+\sqrt{\mu}}_{Q(T)}$ and $\ket{+\sqrt{\mu}}_{Q'(T')}$; see Appendix \ref{coupler} (Eq. \eqref{gamma}) for reference. The quantity $P_{\text{nn}}$ is the probability of $\varkappa=-$ to occur given both \textit{Alice} and \textit{Bob} do not send coherent states in a $Z$ (signal) round. It is given by
\begin{equation}
	\begin{split}
		P_{\text{nn}} &= p_{\text{dark}}(1-p_{\text{dark}}).
	\end{split}
\end{equation}
This probability thus corresponds to an occurrence of a dark count at one detector and a lack of a dark count at the other detector. \\
\indent According to \eqref{error-corr}, $\delta^{-,Z,Z}_{\text{EC}}$ is given by 
\begin{equation}
	\delta^{-,Z,Z}_{\text{EC}} = f_{\text{EC}}H(e^{-,Z,Z}).
\end{equation}
Here $e^{-,Z,Z}$ is evaluated by
\begin{equation}
	\begin{split}
		e^{-,Z,Z} &= e^{-,Z,Z}_1 + e^{-,Z,Z}_2
	\end{split}		
\end{equation}
where
\begin{equation}
	\begin{split}	
		e^{-,Z,Z}_1 &= \frac{\epsilon^2P_{\text{ss}}}{2\epsilon(1-\epsilon)P_{\text{sns}} + \epsilon^2P_{\text{ss}} + (1-\epsilon)^2P_{\text{nn}}}
	\end{split}
\end{equation}
and
\begin{equation}
	\begin{split}	
		e^{-,Z,Z}_2 &= \frac{(1-\epsilon)^2P_{\text{nn}}}{2\epsilon(1-\epsilon)P_{\text{sns}} + \epsilon^2P_{\text{ss}} + (1-\epsilon)^2P_{\text{nn}}}.
	\end{split}
\end{equation}
Thus all the events ($-,Z,Z$) corresponding to configurations (scenarios) \textbf{ss} and \textbf{nn} are regarded as \textit{errors}. Note that $e^{-,Z,Z}_1=p_{k=1,y=1|\varkappa=-}$ and $e^{-,Z,Z}_2=p_{k=0,y=0|\varkappa=-}$. Recall that $k=\overline{y}$ ($k=0,y=1$ or $k=1,y=0$) is required for \textit{Alice} and \textit{Bob} to establish correlated key bits (see \textit{Step 6} in Sec. \ref{protocol}).\\
\indent To determine $\chi(K:E)^{-,Z,Z}$, we use the fact that it reduces to $H\left(1-	\abs{\braket{\Theta^{-,Z,Z}_{0,1}}{\Theta^{-,Z,Z}_{1,0}}}/2\right)$ (see Sec. \ref{loss-scenario}). Under realistic conditions (imperfections), the overlap between \textit{Eve}'s conditional states takes the form  
\begin{widetext}
\begin{equation}
	\begin{split}
		\braket{\Theta^{-,Z,Z}_{0,1}}{\Theta^{-,Z,Z}_{1,0}} = \frac{\bra{+\sqrt{\mu},-\sqrt{\mu}}F^{-}_{\text{imp}}\ket{-\sqrt{\mu},+\sqrt{\mu}}}{\sqrt{\bra{+\sqrt{\mu},-\sqrt{\mu}}F^{-}_{\text{imp}}\ket{+\sqrt{\mu},-\sqrt{\mu}}\bra{-\sqrt{\mu},+\sqrt{\mu}}F^{-}_{\text{imp}}\ket{-\sqrt{\mu},+\sqrt{\mu}}}}
	\end{split}
\end{equation}
\end{widetext}
as shown in Ref.~\cite{lin-lutkenhaus}. \\
\indent In case \textit{active odd-parity pairing} (AOPP)~\cite{aopp,zigzag} is performed, the overall rate of SNS-PM-QKD has the form
\begin{equation}\label{aopp-key-rate}
	\begin{split}
		\tilde{R}_{\text{real}} &= s \times \left(2\epsilon(1-\epsilon)P_{\text{sns}} + \epsilon^2P_{\text{ss}} + (1-\epsilon)^2P_{\text{nn}}\right)\times \\
		&[1 - \chi(K:E)^{-,Z,Z} - \tilde{\delta}^{-,Z,Z}_{\text{EC}}].
	\end{split}
\end{equation}
where 
\begin{equation}
	\begin{split}
		s = \frac{1}{2}[(1-e^{-,Z,Z})^2 + (e^{-,Z,Z})^2]
	\end{split}
\end{equation}
is the \textit{sifting factor} of AOPP, and
\begin{equation}
	\begin{split}
		\tilde{\delta}^{-,Z,Z}_{\text{EC}} = f_{\text{EC}}H(\tilde{e}^{-,Z,Z})
	\end{split}
\end{equation}
is the average amount of information leakage during the error correction procedure in the presence of AOPP algorithm. Take into account that
\begin{equation}
	\begin{split}
		\tilde{e}^{-,Z,Z} = \frac{(e^{-,Z,Z})^2}{(1-e^{-,Z,Z})^2 + (e^{-,Z,Z})^2}
	\end{split}
\end{equation}
is the error rate mitigated by the AOPP procedure. \\
\indent The SNS-PM-QKD with phase randomization (Sec.~\ref{sns-pm-qkd-r}) has an overall key rate of the form
\begin{equation}\label{rand-key-rate}
	\begin{split}
		R'_{\text{real}} &= \frac{1}{2} \left(2\epsilon(1-\epsilon)P_{\text{sns}} + \epsilon^2P'_{\text{ss}} + (1-\epsilon)^2P_{\text{nn}}\right)\times\\
		&[1 - \chi(K:E)^{-,Z,Z} - \tilde{\delta}^{-,Z,Z}_{\text{EC}}]
	\end{split}
\end{equation}
where the fraction $1/2$ identifies the \textit{phase-interval sifting factor}. The quantity $P'_{\text{ss}}$ is evaluated by
\begin{equation}
	\begin{split}
		P'_{\text{ss}} &= \bra{\sqrt{\psi},\sqrt{\psi}}F^{-}_{\text{imp}}\ket{\sqrt{\psi},\sqrt{\psi}}
	\end{split}
\end{equation}
where $\ket{\sqrt{\psi}}$ signifies the output state of a coupler in case of phase mismatch $\delta$ between input states $\ket{\mathrm{e}^{i\nu_a}\sqrt{\mu}}_{Q(T)}$ and $\ket{\mathrm{e}^{i\nu_b}\sqrt{\mu}}_{Q'(T')}$; see Eq.~\eqref{psi} in Appendix~\ref{coupler} for reference. Hence
\begin{equation}\label{rand-aopp-key-rate}
	\begin{split}
		\tilde{R}'_{\text{real}} &= \frac{1}{2}s\times \left(2\epsilon(1-\epsilon)P_{\text{sns}} + \epsilon^2P'_{\text{ss}} + (1-\epsilon)^2P_{\text{nn}}\right)\times\\
		&[1 - \chi(K:E)^{-,Z,Z} - \tilde{\delta}^{-,Z,Z}_{\text{EC}}]
	\end{split}
\end{equation}
is the overall key rate of a SNS-PM-QKD with randomization and AOPP procedure.\\
\indent Results concerning rate-distance functions $R_{\text{real}}(L)$, $R'_{\text{real}}(L)$, $\tilde{R}_{\text{real}}(L)$, and $\tilde{R}'_{\text{real}}(L)$ are given in Sec. \ref{results}. These functions are used to evaluate the contribution of this work (SNS-PM-QKD and SNS-PM-QKD with randomization) compared to existing protocols of the same kind.

\subsection{Results}\label{results}
\begin{figure}[!hbt]
	\centering
	\includegraphics[scale=0.75]{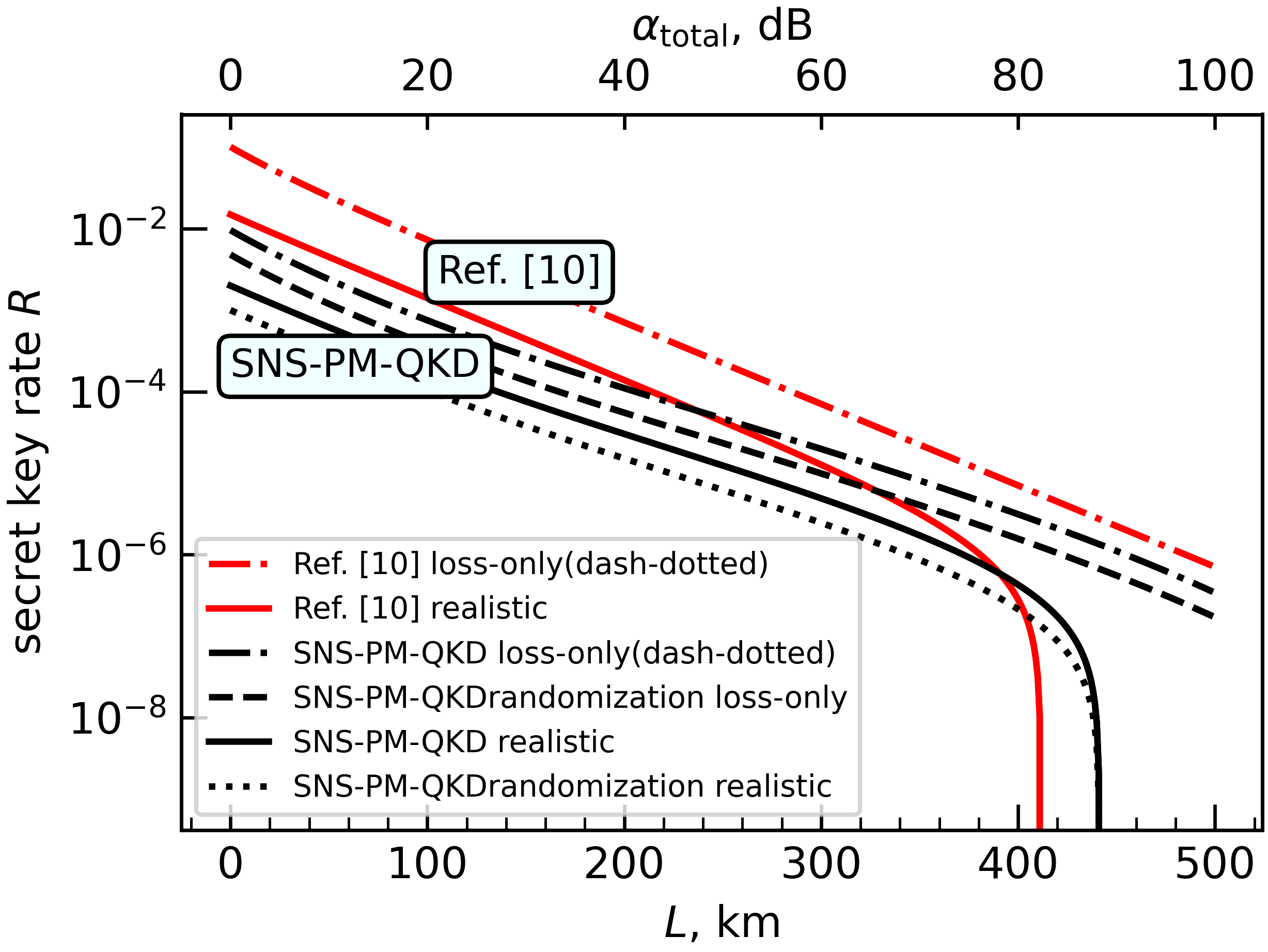}
	\caption{\label{comparison-1} Key-rate comparison of SNS-PM-QKD ($R_{\text{real}}(L)$ with black solid line, $R'_{\text{real}}(L)$ with black dotted line, $R(L)$ with black dash-dotted line, and $R'(L)$ with black dashed line) to PM-QKD of Ref.~\cite{lin-lutkenhaus} (loss-only rate with red dash-dotted line and realistic rate with red solid line). The comparison involves both loss-only ($R(L)$ and $R'(L)$) and realistic ($R_{\text{real}}(L)$ and $R'_{\text{real}}(L)$) scenarios. The functions $R(L)$ and $R'(L)$ are evaluated for $\mu=0.1$ and distance-dependent sending probability $\epsilon(L)$ (see Fig. \ref{epsilon-profile}) by using Eq. \eqref{loss-key-rate} and Eq. \eqref{loss-key-rate-r}, respectively.		
	For the realistic scenario, the rate-distance plot of SNS-PM-QKD (SNS-PM-QKD with randomization) is evaluated using Eq.~\ref{real-key-rate} (Eq.~\ref{rand-key-rate}) for $\mu = 0.1$ and a distance-dependent sending probability $\epsilon(L|\epsilon_0=0.05,\epsilon_{\text{max}}=0.45,L_{\text{max}}=450\text{km})$ of the form shown in Fig. \ref{epsilon-profile}. To ensure consistency with Ref.~\cite{lin-lutkenhaus}, the key rates in the realistic scenario are evaluated for protocol parameters $p_{\text{dark}}=8\times10^{-8}$, $\delta=\frac{\pi}{60}$, $V=0.95$, $\eta_{\text{det}}=0.145$, $f_{\text{EC}} = 1.15$. For both loss-only and realistic scenarios, we assume an attenuation of $\alpha=0.2$ dB/km. In the realistic scenario, SNS-PM-QKD (SNS-PM-QKD with randomization) has a distancerange of $441$km.} 
		\end{figure}

\begin{figure}[!hbt]
\centering
\includegraphics[scale=0.75]{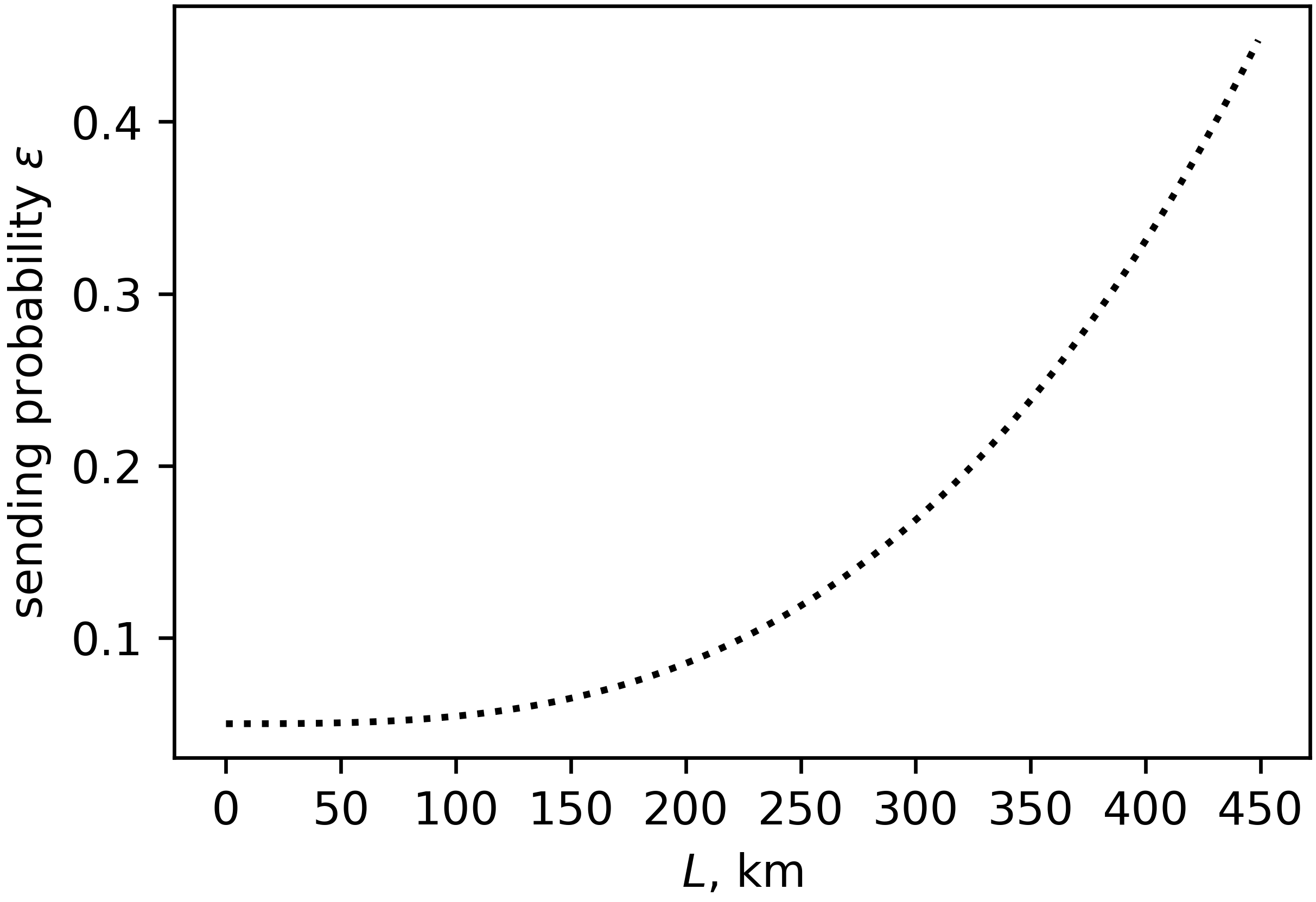}
\caption{\label{epsilon-profile} $\epsilon(L)$ profile used for evaluating the key rate of SNS-PM-QKD. Note that $\epsilon(L)=\epsilon_0+(\epsilon_{\text{max}}-\epsilon_0)\times(L/L_{\text{max}})^3$ where we choose $\epsilon_0=0.05$, $\epsilon_{\text{max}}=0.45$ so that $\epsilon(L)$ is consistent with the approach used in ~\cite{sns-post}\textemdash usage of varying sending probability $\epsilon$ throughout the distance range of the protocol. The value of $L_{\text{max}}$ is conditioned on the comparison: \textbf{(i)} $L_{\text{max}}=450$km; \textbf{(ii)} $L_{\text{max}}=950$km; \textbf{(iii)} $L_{\text{max}}=1100$km. The value of $L_{\text{max}}$ is chosen so that the profile $\epsilon(L)$ is spread over the rate-distance range of SNS-PM-QKD given $\delta\rightarrow0$. In the plot above, we present $\epsilon(L)$ for $L_{\text{max}}=450$km.}
\end{figure}

\begin{figure}[!hbt]
\centering
\includegraphics[scale=0.75]{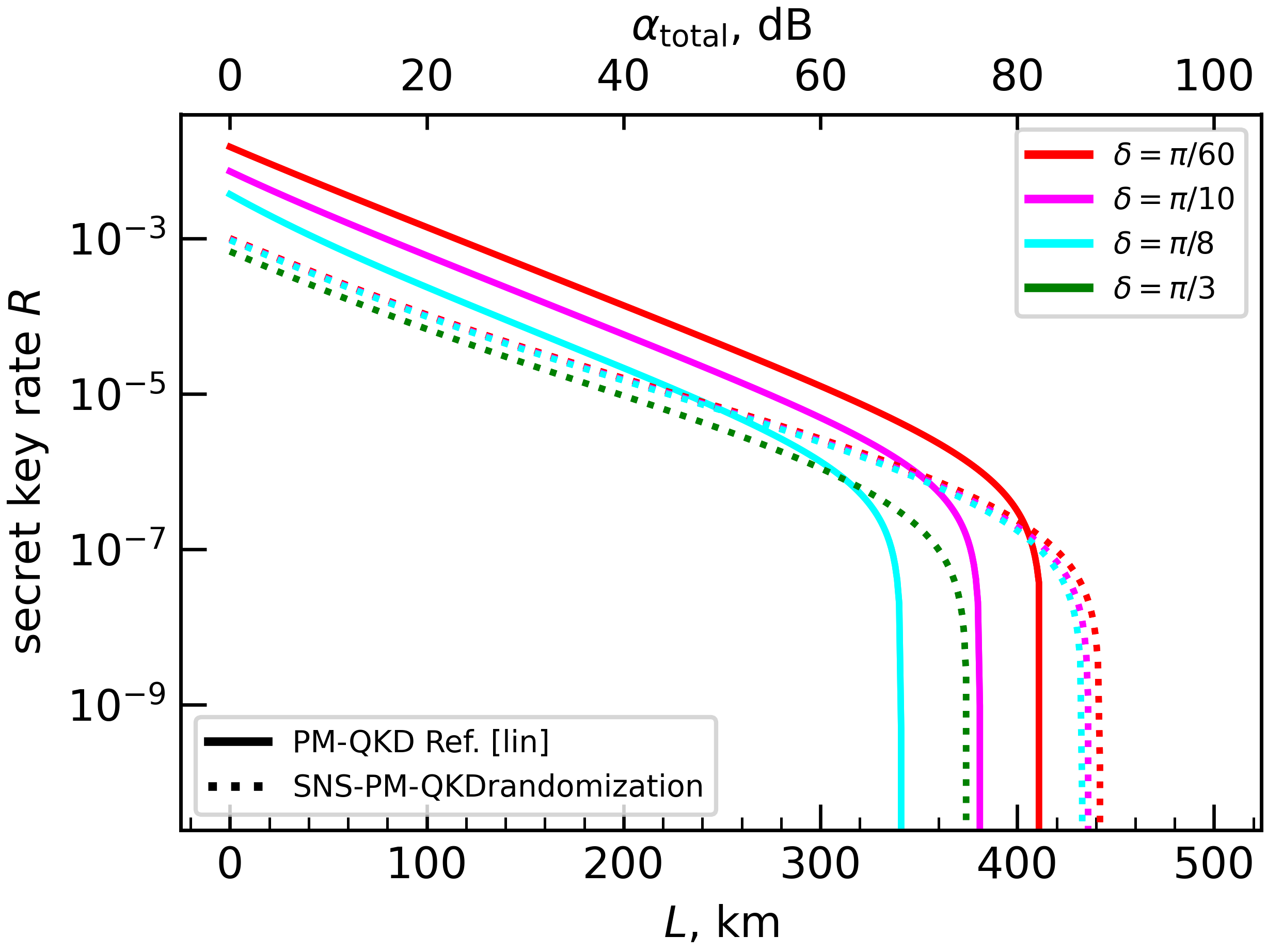}
\caption{\label{comparison-1-1} Error-tolerance comparison of SNS-PM-QKD with randomization ($R'(L)$, Eq. \eqref{rand-key-rate}) to PM-QKD of Ref.~\cite{lin-lutkenhaus}. To ensure a proper error-tolerance comparison, the rate-distance plots of SNS-PM-QKD with randomization and PM-QKD of Ref. \cite{lin-lutkenhaus} are evaluated for protocol parameters $\mu=0.1$, distance-dependent sending probability $\epsilon(L|\epsilon_0=0.05,\epsilon_{\text{max}}=0.45,L_{\text{max}}=450\text{km})$ (see Fig. \ref{epsilon-profile}), $p_{\text{dark}}=8\times10^{-8}$, $V=0.95$, $\eta_{\text{det}}=0.145$, $f_{\text{EC}} = 1.15$, and several values of $\delta$. As shown in the figure, four values of $\delta$ are considered: $\delta=\pi/60$, $\delta=\pi/10$, $\delta=\pi/8$, and  $\delta=\pi/3$.}
\end{figure}

\begin{figure}[!hbt]
\centering
\includegraphics[scale=0.75]{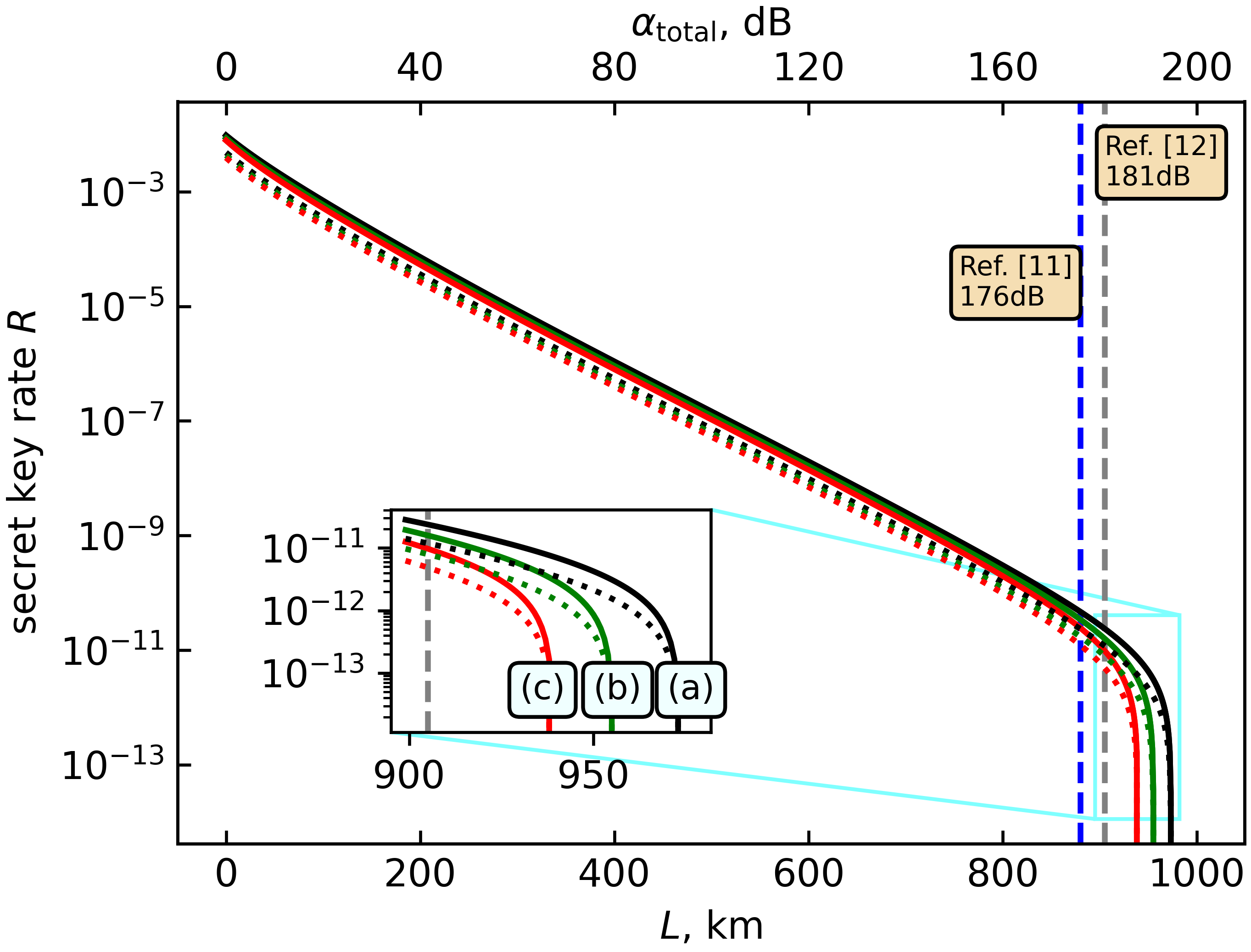}
\caption{\label{comparison-2} Range comparison of SNS-PM-QKD ($\tilde{R}_{\text{real}}(L)$ with solid line, $\tilde{R}'_{\text{real}}(L)$ with dotted line) to both SNS-TF-QKD ~\cite{sns} and SNS-TF-QKD with phase postselection ~\cite{sns-post}. Note that active odd-parity pairing (AOPP) is taken into consideration in the comparison. The key rates $\tilde{R}_{\text{real}}(L)$ and $\tilde{R}'_{\text{real}}(L)$ of SNS-PM-QKD are evaluated by Eq. \eqref{aopp-key-rate} and Eq. \eqref{rand-aopp-key-rate}, respectively, for  $\epsilon=f(L)$ given in Fig. \ref{epsilon-profile}, where $L_{\text{max}}=900$km. According to the results presented in ~\cite{sns-post}, the maximum achievable distance of SNS-TF-QKD is $L=880$km (corresponding to $\alpha_{\text{total}}=176$dB) and that of SNS-TF-QKD with phase postselection is $L=905$km (corresponding to $\alpha_{\text{total}}=181$dB).
To be consistent with the results of Ref.~\cite{sns-post}, the SNS-PM-QKD key rates are evaluated for protocol parameters $p_{\text{dark}}=10^{-11}$, $\eta_{\text{det}}=1$, $\alpha=0.2$dB/km, $f_{\text{EC}} = 1.1$. The intensity parameter $\mu$ is chosen to be $0.1$ so that SNS-PM-QKD achieves maximum distance. Note that the misalignment in SNS-PM-QKD is modeled differently from that of Ref.~\cite{sns-post}. To provide a proper range comparison, we present SNS-PM-QKD rate-distance plots for several values of the pair ($\delta$,$V$): \textbf{(a)} ($\delta=\frac{\pi}{60}$,$V=0.95$); \textbf{(b)} ($\delta=\frac{\pi}{4}$,$V=0.9$); \textbf{(c)} ($\delta=\frac{\pi}{3}$,$V=0.85$). The key rates of the SNS-PM-QKD are obtained for $\mu=0.1$. SNS-PM-QKD (SNS-PM-QKD with randomization) has a distance range of $973$km for \textbf{(a)}, $955$km for \textbf{(b)}, and $938$km for \textbf{(c)}. }
	\end{figure}

\begin{figure}[!hbt]
\centering
\includegraphics[scale=0.75]{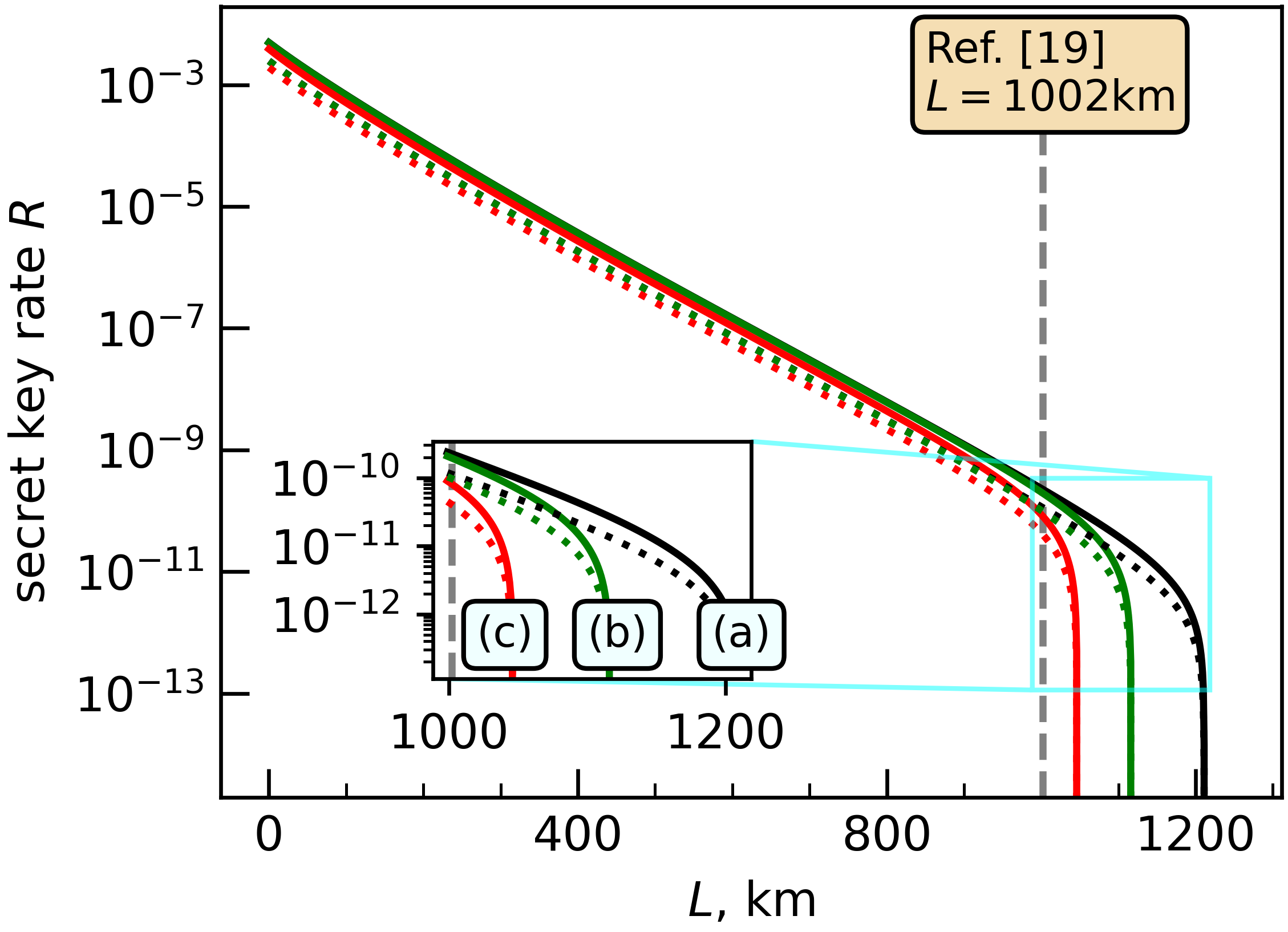}
\caption{\label{comparison-3} Range comparison of SNS-PM-QKD ($\tilde{R}_{\text{real}}(L)$ with solid line, $\tilde{R}'_{\text{real}}(L)$ with dotted line) to experimental SNS-TF-QKD of Ref.~\cite{sns-exp}. Note that active odd-parity pairing (AOPP) is taken into consideration in the comparison. The key rates of SNS-PM-QKD are evaluated by Eq. \eqref{aopp-key-rate} for $\epsilon=f(L)$ given in Fig. \ref{epsilon-profile}, where $L_{\text{max}}=1200$km. According to the results presented in ~\cite{sns-exp}, the maximum achievable distance of the experimental SNS-TF-QKD is $L=1002$km. To be consistent with results of Ref.~\cite{sns-exp}, the SNS-PM-QKD key rates are evaluated for protocol parameters $\eta_{\text{det}}=0.6$, $\alpha_{\text{average}}=0.157$dB/km, $f_{\text{EC}}=1.16$. The intensity parameter $\mu$ is chosen to be $0.1$ so that SNS-PM-QKD achieves maximum distance. Note that both misalignment and dark count rate in SNS-PM-QKD are modeled differently from those of Ref.~\cite{sns-exp}. To provide a proper range comparison, we present SNS-PM-QKD rate-distance plots for several values of the triplet ($\delta$,$V$,$p_{\text{dark}}$): \textbf{(a)} ($\delta=\frac{\pi}{60}$,$V=0.95$,$p_{\text{dark}}=10^{-11}$); \textbf{(b)} ($\delta=\frac{\pi}{35}$,$V=0.9$,$p_{\text{dark}}=5\times10^{-11}$); \textbf{(c)} ($\delta=\frac{\pi}{3}$,$V=0.85$,$p_{\text{dark}}=10^{-10}$). SNS-PM-QKD (SNS-PM-QKD with randomization) has a distance range of $1211$km for \textbf{(a)}, $1116$km for \textbf{(b)}, and $1046$km for \textbf{(c)}.  }
\end{figure}

\indent To objectively evaluate the contribution of this work, we present the following comparisons: \textbf{(i)} range and error-tolerance comparisons of SNS-PM-QKD (and SNS-PM-QKD with randomization) to PM-QKD of Ref.~\cite{lin-lutkenhaus}; \textbf{(ii)} range comparison of SNS-PM-QKD (and SNS-PM-QKD with randomization) to both SNS-TF-QKD of Ref.~\cite{sns} and SNS-TF-QKD with phase postselection ~\cite{sns-post}; \textbf{(iii)} range comaprison of SNS-PM-QKD (and SNS-PM-QKD with randomization) to experimental SNS-TF-QKD of Ref.~\cite{sns-exp}. \\
\textit{Comparison} \textbf{(i)}. This comparison (Fig. \ref{comparison-1}) is used to show that the proposed protocols (SNS-PM-QKD and SNS-PM-QKD with randomization) provide improved range performance over PM-QKD of Ref.~\cite{lin-lutkenhaus}. Both loss-only and realistic scenarios are considered. In the loss-only scenario, the key rate of SNS-PM-QKD (SNS-PM-QKD with randomization) is evaluated by Eq. \eqref{loss-key-rate} (Eq. \eqref{loss-key-rate-r}). To provide a proper comparison in the realistic scenario, the key rates of SNS-PM-QKD (Eq. \eqref{real-key-rate} and Eq. \eqref{rand-key-rate}) and PM-QKD are evaluated with respect to identical protocol parameters: $p_{\text{dark}}=8\times10^{-8}$, $\delta=\frac{\pi}{60}$, $V=0.95$, $\eta_{\text{det}}=0.145$, $f_{\text{EC}} = 1.15$. Take into account that $\mu$ (intensities of both signal and secondary coherent states) takes the optimal value of $0.1$ in SNS-PM-QKD, $\mu=0.1$. To be consistent with the nature of \textit{sending-or-not-sending} approach ~\cite{sns,sns-post}, in SNS-PM-QKD, we adopt a sending probability $\epsilon(L)$ (a function of the transmission distance $L$) of the form shown in Fig. \ref{epsilon-profile}. The function $\epsilon(L)$ is expressed by
\begin{equation}\label{epsilon-eq}
	\epsilon(L) = \epsilon_0+(\epsilon_{\text{max}}-\epsilon_0)\left(\frac{L}{L_{\text{max}}}\right)^3
\end{equation}
where $\epsilon_0=0.05$ and $\epsilon_{\text{max}}=0.45$. The value of $L_{\text{max}}$ is conditioned on the rate-distance range of SNS-PM-QKD in case of $\delta\rightarrow0$\textemdash $\epsilon(L)$ needs to be spread over the rate-distance range. In this comparison, $L_{\text{max}}=450$km.\\
\indent As shown in Fig. \ref{comparison-1}, both SNS-PM-QKD and SNS-PM-QKD with randomization achieve a greater rate-distance range compared to PM-QKD of Ref. \cite{lin-lutkenhaus}. Both proposed protocols reach distances of up to $442$km in this comparison. Two reasons underlie the greater rate-distance range of SNS-PM-QKD (SNS-PM-QKD with randomization) compared to PM-QKD of \cite{lin-lutkenhaus}: \textit{Reason} \textbf{1}: SNS-PM-QKD (SNS-PM-QKD with randomization) is characterized by lower error rate than PM-QKD of Ref.~\cite{lin-lutkenhaus}; \textit{Reason} \textbf{2}: SNS-PM-QKD (SNS-PM-QKD with randomization) exhibits a greater key-rate\textendash to\textendash error-rate ratio (key rate : error rate) than PM-QKD of Ref.~\cite{lin-lutkenhaus}. SNS-PM-QKD has a lower error rate due to \textit{outcome sifting} and low probability of obtaining a key-generation POVM outcome ($\varkappa=-$) given \textbf{ss} and \textbf{nn} scenarios. SNS-PM-QKD exhibits a higher key-rate\textendash to\textendash error-rate ratio due to \textit{Reason} \textbf{1} and the use of a distance-dependent sending probability $\epsilon(L)$ (taking values of up to approximately 0.45) that yields the corresponding rate-distance range (Fig. \ref{comparison-1}).  \\
\indent Furthermore \textit{comparison} \textbf{(i)} is used to show that the proposed protocols (SNS-PM-QKD and SNS-PM-QKD with randomization) exhibit improved error tolerance compared to the PM-QKD protocol of Ref.~\cite{lin-lutkenhaus} (see Fig. \ref{comparison-1-1}). The error-tolerance comparison is performed by analyzing the behavior of SNS-PM-QKD with randomization and PM-QKD for several values of the phase mismatch $\delta$. Specifically four values are considered: $\delta=\pi/60$, $\delta=\pi/10$, $\delta=\pi/8$, $\delta=\pi/3$. As shown in Fig. \ref{comparison-1-1}, SNS-PM-QKD with randomization exhibits a relatively high tolerance to phase mismatch (misalignment) tolerance, significantly exceeding that of PM-QKD. As $\delta$ increases, the achievable distance of SNS-PM-QKD with randomization deteriorates at a slower rate than that of PM-QKD. Moreover, for $\delta=\pi/3$, PM-QKD fails to yield a positive key rate; accordingly, the corresponding rate-distance curve is absent from the figure.  \\ 
\textit{Comparison} \textbf{(ii)}. This comparison (Fig. \ref{comparison-2}) is used to show that the proposed protocols (SNS-PM-QKD with or without randomization) provide improved range performance over SNS-TF-QKD ~\cite{sns} and SNS-TF-QKD with postselection ~\cite{sns-post}. We are interested only in the maximum achievable distances of SNS-TF-QKD and SNS-TF-QKD with postselection: $176$dB and $181$dB, respectively. Active odd-parity pairing (AOPP) procedure is involved in the performances of the protocols in this comparison. In Fig. \ref{comparison-2}, SNS-PM-QKD SNS-PM-QKD with randomization are represented by rate-distance functions $\tilde{R}_{\text{real}}(L)$ (see Eq. \eqref{aopp-key-rate}) and $\tilde{R}'_{\text{real}}(L)$ (see Eq. \eqref{rand-aopp-key-rate}), respectively. For these rate-distance functions,  we use sending probability profile $\epsilon(L)$ with $\epsilon_0=0.05$, $\epsilon_{\text{max}}=0.45$, and $L_{\text{max}}=950$km (see Fig. \ref{epsilon-profile} and Eq. \eqref{epsilon-eq}). To provide a proper comparison, we evaluate the key rates of SNS-PM-QKD (SNS-PM-QKD with randomization) for protocol parameters $p_{\text{dark}}=10^{-11}$, $\eta_{\text{det}}=1$, $\alpha=0.2$dB/km, $f_{\text{EC}}=1.1$. These values of the protocol parameters are adopted from Ref.~\cite{sns-post}. Also we present the functions $R'_{\text{real}}(L)$ and $\tilde{R}'_{\text{real}}(L)$ for three scenarios: ($\delta=\frac{\pi}{60}$,$V=0.95$); ($\delta=\frac{\pi}{60}$,$V=0.95$); ($\delta=\frac{\pi}{60}$,$V=0.95$). Take into account that $\mu$ (intensities of both signal and secondary coherent states) has the value of $0.1$ in SNS-PM-QKD (SNS-PM-QKD with randomization).  \\
\indent As shown in Fig. \ref{comparison-2}, both SNS-PM-QKD and SNS-PM-QKD with randomization exceed the maximum achievable distances of SNS-TF-QKD \cite{sns} and SNS-TF-QKD with postselection \cite{sns-post}\textemdash this holds for all the ($\delta$,$V$) scenarios (\textbf{(a)}, \textbf{(b)}, and \textbf{(c)}) considered in this comparison. Given the protocol parameters above, SNS-PM-QKD (SNS-PM-QKD with randomization) achieves a maximum distance of $973$km.\\
\textit{Comparison} \textbf{(iii)}. This comparison (Fig. \ref{comparison-3}) is used to show that the proposed protocol (SNS-PM-QKD) provides improved range performance over the experimental implementation of SNS-TF-QKD introduced in ~\cite{sns-exp}. We are interested only in the maximum achievable distance of the experimental SNS-TF-QKD: $1002$km. Active odd-parity pairing (AOPP) procedure is involved in the performances of the protocols in this comparison as well. In Fig. \ref{comparison-3}, SNS-PM-QKD and SNS-PM-QKD with randomization are represented by rate-distance functions $\tilde{R}_{\text{real}}(L)$ (see Eq. \eqref{aopp-key-rate}) and $\tilde{R}'_{\text{real}}(L)$ (see Eq. \eqref{rand-aopp-key-rate}), respectively. For these rate-distance functions,  we use sending probability profile $\epsilon(L)$ with $\epsilon_0=0.05$, $\epsilon_{\text{max}}=0.45$, and $L_{\text{max}}=1200$km (see Fig. \ref{epsilon-profile} and Eq. \eqref{epsilon-eq}). To provide a proper comparison, we evaluate the key rates of SNS-PM-QKD (SNS-PM-QKD with randomization) for protocol parameters $\eta_{\text{det}}=0.6$, $\alpha=0.157$dB/km, $f_{\text{EC}}=1.16$. These values of the protocol parameters are adopted from Ref.~\cite{sns-exp}. Also we present the functions $\tilde{R}_{\text{real}}(L)$ and $\tilde{R}'_{\text{real}}(L)$ for three scenarios: ($\delta=\frac{\pi}{60}$,$V=0.95$,$p_{\text{dark}}=10^{-11}$); ($\delta=\frac{\pi}{35}$,$V=0.95$,$p_{\text{dark}}=5\times10^{-11}$); ($\delta=\frac{\pi}{3}$,$V=0.95$,$p_{\text{dark}}=10^{-10}$). Take into account that $\mu$ (intensities of both signal and secondary coherent states) has the value of $0.1$ in SNS-PM-QKD (SNS-PM-QKD with randomization). \\
\indent As shown in Fig. \ref{comparison-3}, both SNS-PM-QKD and SNS-PM-QKD with randomization exceed the rate-distance barrier of the experimental SNS-TF-QKD of Ref. \cite{sns-exp}\textemdash this holds for all the ($\delta$,$V$,$p_{\text{dark}}$) scenarios (\textbf{(a)}, \textbf{(b)}, and \textbf{(c)}) considered in this comparison. Given the protocol parameters above, SNS-PM-QKD (SNS-PM-QKD with randomization) achieves a maximum distance range of $1211$km.\\

\section{Summary}\label{summ}
\indent We have reported a PM-QKD protocol based on the \textit{sending-or-not-sending} approach, referred to as SNS-PM-QKD. We have evaluated its asymptotic key rate under collective attacks using the framework introduced in Ref.~\cite{lin-lutkenhaus}. We have also analyzed the behavior of the proposed scheme under a double-POVM attack, which constitutes a protocol-specific form of collective attack. The double-POVM attack is verified to be detectable, since it results in a higher error rate than that predicted for the general collective attack within the framework of Ref.~\cite{lin-lutkenhaus}. The rate–distance performance of SNS-PM-QKD is compared with that of PM-QKD \cite{lin-lutkenhaus}, SNS-TF-QKD, SNS-TF-QKD with postselection \cite{sns-post}, and experimental SNS-TF-QKD implementations \cite{sns-exp}, in terms of the maximum achievable transmission distance. SNS-PM-QKD exhibits a relatively high tolerance to a phase mismatch (misalignment errors), significantly exceeding that of PM-QKD. It also achieves greater operational distances than all the other protocols involved in the comparison. Figures~\ref{comparison-2} and \ref{comparison-3} demonstrate that the proposed protocol surpasses the current rate-distance performance of both theoretical SNS-TF-QKD \cite{sns-post} and the experimental SNS-TF-QKD implementation of Ref.~ \cite{sns-exp}. An interesting direction for future work is to investigate the performance of SNS-PM-QKD in the finite-key regime. It would also be of interest to examine the achievable rate–distance behavior in experimental implementations of SNS-PM-QKD.

\begin{acknowledgments}
	The work has been performed at Technical University of Varna, Bulgaria.
	This work was supported by Technical University of Varna under Scientific Project Programme, Grant No. H$\Pi$6/2025. 
\end{acknowledgments}

\appendix
\section{Coupler operation}\label{coupler}
\indent A coupler consists of two inputs and a single output. A destructive interference occurs at the output, \textit{i.e.}, the state $\ket{T(\mathrm{e}^{i\varphi}\alpha) - R(\mathrm{e}^{i\phi}\beta)}$ appears. Note that $\alpha$ and $\beta$ are the intensities (amplitudes) of the input coherent states whereas $\varphi$ and $\phi$ are their respective phases. Also $T$ and $R$ are \textit{transmittance} and \textit{reflectance} of the coupler, respectively. We assume that $T = R = 1$. The above output state represents an ideal scenario, where there is neither \textit{phase mismatch} nor \textit{mode} (\textit{intensity}) \textit{mismatch} between the input states. If phase mismatch is considered, a phase shift needs to be induced into one of the input states. In this regard, we assume that the phase mismatch angle $\delta$ is induced into the reflected state, \textit{i.e.}, the output state takes the form $\ket{\mathrm{e}^{i\varphi}\alpha - \mathrm{e}^{i\delta}\mathrm{e}^{i\phi}\beta}$ [\textit{Note}: The parameter $\delta$ could be considered as constant ~\cite{lin-lutkenhaus}]. If mode (intensity) mismatch is considered, we induce asymmetry into the input intensities \textemdash $\ket{\mathrm{e}^{i\varphi}\alpha - \mathrm{e}^{i\delta}\mathrm{e}^{i\phi}\sqrt{V}\beta}$, where $V<1$ in the non-ideal case.\\
\indent For the sake of clarity, we work out a simple example below. Consider a scenario in which $\alpha = \beta = \sqrt{\mu}$, $\varphi = \phi = 0$, $\delta = \frac{\pi}{60}$, $V=0.95$\textemdash this scenario corresponds to \textbf{ss} configuration and $Z$ (signal) round chosen by both \textit{Alice} and \textit{Bob} in the SNS-PM-QKD protocol (see Sec. \ref{sns-pm-qkd}). A coupler's output state in general takes the form 
\begin{equation*}
	\begin{split}
		\ket{\mathrm{e}^{i\varphi}\sqrt{\mu} - \mathrm{e}^{i\delta}\mathrm{e}^{i\phi}\sqrt{V}\sqrt{\mu}} &=  \ket{\mathrm{e}^{i\phi}(\mathrm{e}^{i(\varphi-\phi)}-\mathrm{e}^{i\delta}\sqrt{V})\sqrt{\mu}}\\
		&= \ket{\mathrm{e}^{i\phi}(\mathrm{e}^{i\Delta}-\mathrm{e}^{i\delta}\sqrt{V})\sqrt{\mu}} = \ket{\sigma}. 
	\end{split}
\end{equation*}
This state can be rewritten as 
\begin{equation*}
	\ket{\abs{\sigma}\cdot \mathrm{e}^{i\omega}}
\end{equation*}
where
\begin{equation*}
	\begin{split}
		\abs{\sigma} &= \sqrt{\Re(\sigma)^2+\Im(\sigma)^2} = \sqrt{(1+V-2\sqrt{V}\cos(\Delta-\delta))\mu},\\ \omega&=\atan(\frac{\Im(\sigma)}{\Re(\sigma)}).
	\end{split}
\end{equation*}
Given $\varphi=\phi=0$, $\delta = \frac{\pi}{60}$ and $V=0.95$, the coupler's output state becomes
\begin{equation*}
	\begin{split}
		\ket{\sqrt{0.00427\mu}\mathrm{e}^{i\omega}} = \ket{\sqrt{\gamma}\mathrm{e}^{i\omega}}.
	\end{split}
\end{equation*}
We now express the above coherent state in a canonical basis, following Ref.~\cite{lin-lutkenhaus}. A single coherent state
\begin{equation}
	\ket{\sqrt{\gamma}\mathrm{e}^{i\omega}} = \mathrm{e}^{-\frac{1}{2}|\sqrt{\gamma}\mathrm{e}^{i\omega}|^2}\sum_{n=0}^{\infty}\frac{(\sqrt{\gamma}\mathrm{e}^{i\omega})^n}{\sqrt{n!}}\ket{n},
\end{equation}
can be expressed in terms of the basis $\{\ket{e_0},\ket{e_1}\}$~\cite{lin-lutkenhaus}
\begin{equation}\label{basis}
	\ket{\sqrt{\gamma}\mathrm{e}^{i\omega}} = c_0(\mathrm{e}^{i\omega})^{2n}\ket{e_0} + c_1(\mathrm{e}^{i\omega})^{2n+1}\ket{e_1},
\end{equation}
where
\begin{eqnarray}\label{basis-vectors}
	\begin{split}
		&\ket{e_0} = \frac{1}{\sqrt{\cosh(|\sqrt{\gamma}\mathrm{e}^{i\omega}|^2)}}\sum_{n=0}^{\infty}\frac{(\sqrt{\gamma})^{2n}}{\sqrt{2n!}}\ket{2n}, \\
		&\ket{e_1} = \frac{1}{\sqrt{\sinh(|\sqrt{\gamma}\mathrm{e}^{i\omega}|^2)}}\sum_{n=0}^{\infty}\frac{(\sqrt{\gamma})^{2n+1}}{\sqrt{(2n+1)!}}\ket{2n+1},
	\end{split}
\end{eqnarray}
and
\begin{eqnarray}\label{coeff-ss}
	\begin{split}
		&c_0(\gamma) = \mathrm{e}^{-\frac{|\sqrt{\gamma}\mathrm{e}^{i\omega}|^2}{2}}\sqrt{\cosh(|\sqrt{\gamma}\mathrm{e}^{i\omega}|^2)}, \\
		&c_1(\gamma) = \mathrm{e}^{-\frac{|\sqrt{\gamma}\mathrm{e}^{i\omega}|^2}{2}}\sqrt{\sinh(|\sqrt{\gamma}\mathrm{e}^{i\omega}|^2)}.
	\end{split}
\end{eqnarray}
As noted in Ref.~\cite{lin-lutkenhaus}, a complex factor $\mathrm{e}^{i\omega}$ (coefficients $(\mathrm{e}^{i\omega})^{2n}$ and $(\mathrm{e}^{i\omega})^{2n+1}$ in Eq. \eqref{basis}) can be absorbed into the definitions of $\ket{e_0}$ and $\ket{e_1}$. That is, Eq. \eqref{basis-vectors} becomes
\begin{eqnarray}\label{basis-vectors-comp}
	\begin{split}
		&\ket{e_0'} = \frac{(\mathrm{e}^{i\omega})^{2n}}{\sqrt{\cosh(|\sqrt{\gamma}\mathrm{e}^{i\omega}|^2)}}\sum_{n=0}^{\infty}\frac{(\sqrt{\gamma})^{2n}}{\sqrt{2n!}}\ket{2n}, \\
		&\ket{e_1'} = \frac{(\mathrm{e}^{i\omega})^{2n+1}}{\sqrt{\sinh(|\sqrt{\gamma}\mathrm{e}^{i\omega}|^2)}}\sum_{n=0}^{\infty}\frac{(\sqrt{\gamma})^{2n+1}}{\sqrt{(2n+1)!}}\ket{2n+1}.
	\end{split}
\end{eqnarray}
Therefore, in the case of a \textbf{ss} configuration, the output state of a coupler can be expressed in a canonical form with real-valued amplitudes as follows
\begin{equation}\label{gamma}
	\ket{\sqrt{\gamma}\cdot \mathrm{e}^{i\omega}} \rightarrow \ket{\sqrt{\gamma}} = c_0(\gamma)\ket{e_0'} - c_1(\gamma)\ket{e_1'}.
\end{equation}
\indent In the case of \textbf{ss} configuration given SNS-PM-QKD with phase randomization (see Sec. \ref{sns-pm-qkd-r}), a coupler's output state takes the form
\begin{equation*}
	\ket{\mathrm{e}^{i\nu_b}(\mathrm{e}^{i(\nu_a-\nu_b)}-\mathrm{e}^{i\delta}\sqrt{V})\sqrt{\mu}}=\ket{\abs{\lambda}\cdot \mathrm{e}^{i\zeta}}
\end{equation*}
where
\begin{equation*}
	\begin{split}
		\abs{\lambda} &= \sqrt{(1+V-2\sqrt{V}\cos((\nu_a-\nu_b)-\delta))\mu},\\ \zeta&=\atan(\frac{\Im(\lambda)}{\Re(\lambda)}).
	\end{split}
\end{equation*}	
Note that $-\pi<\nu_a-\nu_b<\pi$ for a given phase interval ($\nu_{a(b)}\in [0,\pi)$ or $\nu_{a(b)}\in[\pi,2\pi)$). Take into account that the highest the intensity $\abs{\lambda}$, the highest the error rate induced by a \textbf{ss} configuration. Hence the worst-case scenario for a \textbf{ss} configuration ($\max(\abs{\lambda})$) is a coupler's output state with intensity
\begin{equation*}
	\begin{split}
		\max\limits_{\nu_a-\nu_b} (\abs{\lambda})&= \sqrt{(1+V-2\sqrt{V}\cos((\delta-\pi)-\delta))\mu} = \sqrt{\psi}.
	\end{split}
\end{equation*}	
This implies that the worst-case scenario for \textbf{ss} configuration in a SNS-PM-QKD with randomization takes place at $\nu_a-\nu_b=\delta-\pi$ (\textit{i.e.}, $\nu_b > \nu_a$ given $\delta<\pi$). The corresponding output coherent state has a canonical form 
\begin{equation}\label{psi}
	\ket{\sqrt{\psi}\cdot \mathrm{e}^{i\zeta}} \rightarrow \ket{\sqrt{\psi}} = c_0(\psi)\ket{e_0'} - c_1(\psi)\ket{e_1'}.
\end{equation}
when one takes into consideration Eqs. \eqref{basis}-\eqref{gamma}. We make use of $\ket{\sqrt{\psi}}$ to evaluate the minimum of the protocol's key rate (Eqs.~\eqref{rand-key-rate} and \eqref{rand-aopp-key-rate})\textemdash in this way, we evaluate the key rate for the worst-case scenario (in the presence of maximum error rate).

\section{POVM operators}\label{povm-operators}
\begin{widetext}
\indent The matrix representations of the POVM operators $F^-$, $F^+$, $F^?$   (loss-only scenario) are given as follows ~\cite{lin-lutkenhaus}
\begin{equation}\label{fMinus}
	\begin{split}
		F^- = (1-\xi^2)\begin{bmatrix}
			\frac{1-\xi^2\Omega^2}{8c_0^4}&\frac{-1+\xi^2\Omega^2}{8c_0^2c_1^2}&0&0 \\
			\frac{-1+\xi^2\Omega^2}{8c_0^2c_1^2}&\frac{1-\xi^2\Omega^2}{8c_1^4}&0&0 \\
			0&0&\frac{1+\xi^2\Omega^2}{8c_0^2c_1^2}&\frac{-1-\xi^2\Omega^2}{8c_0^2c_1^2} \\
			0&0&\frac{-1-\xi^2\Omega^2}{8c_0^2c_1^2}&\frac{1+\xi^2\Omega^2}{8c_0^2c_1^2}
		\end{bmatrix} 
	\end{split}
\end{equation}
\begin{equation}\label{fPlus}
	\begin{split}
		F^+ = (1-\xi^2)\begin{bmatrix}
			\frac{1-\xi^2\Omega^2}{8c_0^4}&\frac{1-\xi^2\Omega^2}{8c_0^2c_1^2}&0&0 \\
			\frac{1-\xi^2\Omega^2}{8c_0^2c_1^2}&\frac{1-\xi^2\Omega^2}{8c_1^4}&0&0 \\
			0&0&\frac{1+\xi^2\Omega^2}{8c_0^2c_1^2}&\frac{1+\xi^2\Omega^2}{8c_0^2c_1^2} \\
			0&0&\frac{1+\xi^2\Omega^2}{8c_0^2c_1^2}&\frac{1+\xi^2\Omega^2}{8c_0^2c_1^2}
		\end{bmatrix} 
	\end{split}
\end{equation}
\begin{equation}\label{fNo}
	\begin{split}
		F^? = \xi^2\begin{bmatrix}
			\frac{(1+\Omega)^2}{4c_0^4}&0&0&0 \\
			0&\frac{(1-\Omega)^2}{4c_1^4}&0&0 \\
			0&0&\frac{1-\Omega^2}{4c_0^2c_1^2}&0 \\
			0&0&0&\frac{1-\Omega^2}{4c_0^2c_1^2}
		\end{bmatrix}
	\end{split}
\end{equation}
The matrix representations of the POVM operators $F^-_{\text{imp}}$, $F^+_{\text{imp}}$ (realistic scenario, imperfections) are
\begin{equation}
	F^-_{\text{imp}} = (1-p_{\text{dark}})F^-_{\text{mis}}+(1-p_{\text{dark}})p_{\text{dark}}F^?_{\text{mis}}
\end{equation}
\begin{equation}
	F^+_{\text{imp}} = (1-p_{\text{dark}})F^+_{\text{mis}}+(1-p_{\text{dark}})p_{\text{dark}}F^?_{\text{mis}}
\end{equation}
where
\begin{widetext}
\begin{equation}
	\begin{split}
		F^-_{\text{mis}} &= \begin{bmatrix}
			\frac{a+b+2c+2d+o+p}{8c_0^4}&-\frac{a+b-o-p}{8c_0^2c_1^2}&0&0 \\
			-\frac{a+b-o-p}{8c_0^2c_1^2}&\frac{a+b-2c-2d+o+p}{8c_1^4}&0&0 \\
			0&0&\frac{a-b+o-p}{8c_0^2c_1^2}&-\frac{a-b+2c-2d-o+p}{8c_0^2c_1^2} \\
			0&0&-\frac{a-b-2c+2d-o+p}{8c_0^2c_1^2}&\frac{a-b+o-p}{8c_0^2c_1^2}
		\end{bmatrix} \\
		F^+_{\text{mis}} &= \begin{bmatrix}
			\frac{a+b+2c+2d+o+p}{8c_0^4}&\frac{a+b-o-p}{8c_0^2c_1^2}&0&0 \\
			\frac{a+b-o-p}{8c_0^2c_1^2}&\frac{a+b-2c-2d+o+p}{8c_1^4}&0&0 \\
			0&0&\frac{a-b+o-p}{8c_0^2c_1^2}&\frac{a-b+2c-2d-o+p}{8c_0^2c_1^2} \\
			0&0&\frac{a-b-2c+2d-o+p}{8c_0^2c_1^2}&\frac{a-b+o-p}{8c_0^2c_1^2}
		\end{bmatrix} \\
		F^?_{\text{mis}} &= \xi^2\begin{bmatrix}
			\frac{(1+\Omega)^2}{4c_0^4}&0&0&0 \\
			0&\frac{(1-\Omega)^2}{4c_1^4}&0&0 \\
			0&0&\frac{1-\Omega^2}{4c_0^2c_1^2}&0 \\
			0&0&0&\frac{1-\Omega^2}{4c_0^2c_1^2}
		\end{bmatrix}
	\end{split}
\end{equation}
\end{widetext}
Note that $\xi=\mathrm{e}^{-\sqrt{\eta}\mu}$, $\Omega=\mathrm{e}^{-2(1-\sqrt{\eta})\mu}$, where $\eta=\eta_{\text{det}}^2\eta_{\text{t}}=\eta_{\text{det}}^210^{-\frac{0.2L}{10}}$ is the \textit{overall} transmittance at distance $L$ [km]. Take into account that ~\cite{lin-lutkenhaus}
\begin{widetext}
\begin{equation}\label{abcdop}
	\begin{split}
		a &= (1-\xi^{1+\sqrt{V}\cos{\delta}})\xi^{1-\sqrt{V}\cos{\delta}}, \;\;\; b = (\xi^{2(1+\sqrt{V}\cos{\delta})}-\xi^{1+\sqrt{V}\cos{\delta}})\xi^{1-\sqrt{V}\cos{\delta}}\Omega^2, \;\;\; c = (\xi^{1+i\sqrt{V}\sin{\delta}}-\xi)\xi\Omega, \\
		d& = (\xi^{1-i\sqrt{V}\sin{\delta}}-\xi)\xi\Omega, \;\;\; o = (1-\xi^{1-\sqrt{V}\cos{\delta}})\xi^{1+\sqrt{V}\cos{\delta}}, \;\;\; p = (\xi^{2(1-\sqrt{V}\cos{\delta})}-\xi^{1-\sqrt{V}\cos{\delta}})\xi^{1+\sqrt{V}\cos{\delta}}\Omega^2
	\end{split}
\end{equation}
\end{widetext}
The coefficients $c_0$ and $c_1$ are defined in Eq.~\ref{coeff-ss}.
\end{widetext}


\bibliography{apssamp}

\end{document}